\documentclass[prb,  twocolumn, showpacs, superscriptaddress,longbibliography]{revtex4-1}
\usepackage{amsmath, amsfonts, amssymb, mathrsfs}
\usepackage{graphicx, subfigure, color}
\usepackage{bm}
\usepackage{braket}
\usepackage{epstopdf}
\usepackage{ulem} 
\usepackage{color}
\usepackage[breaklinks]{hyperref}
\hypersetup{
   colorlinks,
   citecolor=blue,
   filecolor=blue,
   linkcolor=blue,
   urlcolor=blue
}
\graphicspath{Graphes}

\renewcommand{\vec}[1]{\boldsymbol{#1}}

\newcommand{\be}{\begin{equation}}
\newcommand{\ee}{\end{equation}}

\definecolor{orange}{rgb}{1,0.5,0}

\definecolor{grey}{rgb}{.6,.6,.6}

\begin{document}

\title{Many-terminal Majorana island: from Topological to Multi-Channel Kondo Model}

\author{Lo\"{i}c Herviou}
\affiliation{Centre de Physique Th\'{e}orique, \'{E}cole Polytechnique, CNRS, Universit\' e Paris-Saclay, 91128 Palaiseau, France}
\affiliation{Laboratoire Pierre Aigrain, \'Ecole Normale Sup\'erieure-PSL Research University, CNRS, Universit\'e Pierre et Marie Curie-Sorbonne Universit\'es, Universit\'e Paris Diderot-Sorbonne Paris Cit\'e, 24 rue Lhomond, 75231 Paris Cedex 05, France}
\author{Karyn~Le~Hur}
\affiliation{Centre de Physique Th\'{e}orique, \'{E}cole Polytechnique, CNRS, Universit\' e Paris-Saclay, 91128 Palaiseau, France}
\author{Christophe Mora}
\affiliation{Laboratoire Pierre Aigrain, \'Ecole Normale Sup\'erieure-PSL Research University, CNRS, Universit\'e Pierre et Marie Curie-Sorbonne Universit\'es, Universit\'e Paris Diderot-Sorbonne Paris Cit\'e, 24 rue Lhomond, 75231 Paris Cedex 05, France}

\begin{abstract}
We study Kondo screening obtained by coupling Majorana bound states, located on a topological superconducting island, to interacting electronic reservoirs. At the charge degeneracy points of the island, we formulate an exact mapping onto the spin-$1/2$ multi-channel Kondo effect. The coupling to Majorana fermions transforms the tunneling terms into effective fermionic bilinear contributions with a Luttinger parameter $K$ in the leads that is effectively doubled. For strong interaction, $K=1/2$, the intermediate fixed point of the standard multi-channel Kondo model is exactly recovered. It evolves with $K$ and connects to strong coupling in non-interacting case $K=1$, with maximum conductance between the leads and robustness against channel asymmetries similarly to the topological Kondo effect. For a number of leads above four, there exists a window of Luttinger parameters in which a quantum phase transition can occur between the strong coupling topological Kondo state and the partially conducting multi-channel Kondo state.
\end{abstract}

\date{\today}
\maketitle

\section{Introduction}

The realization and manipulation of Majorana bound states in topological superconductors has been recently the focus of numerous studies~\cite{alicea2012,Beenakker2013}. Such exotic quasi-particles appear as many-body fermionic excitations at zero energy, and can be topologically protected from decoherence and other perturbations. Typical examples are the $p+ip$ superconductor in two dimensions~\cite{Read2000} and the Kitaev p-wave superconducting wire~\cite{Kitaev2001}. There have been numerous efforts~\cite{A.Das2012, wire1, wire2} to realize such a one-dimensional p-wave superconductor in InAs quantum wires through a combination of spin-orbit coupling and Zeeman effect. More recently, proposals were made to realize mesoscopic geometries with such wires for further applications in quantum computing~\cite{Albrecht2016}. Theoretically, some recent works have also suggested to observe non-Abelian statistics through gate engineering~\cite{Halperin2011}.

For a normal island without Majorana fermion, Matveev formulated the charge Kondo effect~\cite{matveev1991b,matveev1995} in which two degenerate charge states play the role of an effective spin $1/2$ and hybridize with electrons either in contacted reservoirs or on the island, realizing the multi-channel Kondo model (M-CKM). Remarkably, in the two-channel Kondo model emulated with two leads, an unscreened Majorana excitation appears at low energy~\cite{Emery1992, Clarke1993, Sengupta1994, Coleman1995,Mora2013}. It is however an emergent particle and differs from the proximity-induced Majorana fermions considered in this paper. Recently, the two-channel charge Kondo model has been realized\cite{Iftikhar2015} in a GaAs setting with an unprecedented control over the model parameters. Two-channel Kondo screening has also been observed with a real spin~\cite{potok2007,Keller2015}. In all cases, it requires fine-tuning which makes its experimental characterization challenging. 

Combining charging effects with Majorana fermions, or Kondo physics with Majorana fermions~\cite{golub2011,lee2013,meng2014}, has already been argued to lead to exotic transport dc~\cite{Fu2010, Zazunov2011, vanHeck2016} or ac properties in quantum RC setting~\cite{Golub2012,lee2014}. In this paper, we study a device proposed in the seminal works of Refs. \onlinecite{Beri-2012, Atland-2013}: a single floating (not grounded) Majorana island connecting with $M\ge 3$ reservoir leads, modeled by Luttinger liquids, through separated Majorana zero modes. The "charge", {\it i.e.} number of Cooper pairs plus number of fermions in the zero energy Majorana manifold on the islands, can be varied through a gate voltage coupled to the quantum box. Progress in building such mesoscopic boxes have been made recently\cite{Krogstrup2015, Buccheri2016}. In the non-degenerate case where transport occurs through a single charge state, an unconventional Kondo screening, named "topological Kondo model", has been theoretically explored~\cite{Beri-2012, Atland-2013, Eriksson-2014, Eriksson-2014-Manifold, Beri-2013, Galpin-2014,  Zazunov-2014,  Atland-2014, Plugge-2016} where the SO(M) impurity "spin" is built from the Majorana excitations. The corresponding low-energy theory exhibits non-Fermi liquid exponents captured by a strong coupling quantum Brownian motion (QBM) picture. In this analogy, the effective particle is pinned at the minima of a triangular lattice connected by instantons. Simple expressions for the leading irrelevant operator dimensions can thus be derived in agreement with the more involved Conformal Field Theory approach~\cite{Affleck1993}. 

In contrast with the topological Kondo effect, the multi-channel Kondo model does not admit a simple QBM description. The effective particle moves on a honeycomb lattice and the low-energy fixed point is at intermediate (neither weak or strong) coupling which excludes a full analytical QBM analysis. In addition, this infrared intermediate fixed point is not robust and requires fine-tuning whereas the topological Kondo non-Fermi liquid fixed point is at strong coupling and stable against perturbations such as asymmetric lead couplings. This last point seems to favor the experimental observation of the more robust topological Kondo effect over the standard multi-channel Kondo model.

In this work, we focus on the charge degenerate case where the two charge states $n$ and $n+1$ are energetically equivalent and described by a pseudo-spin. We show that the resulting QBM lattice is triangular with a pseudo-spin texture characterized by a Berry phase. The corresponding dual model is a honeycomb lattice. As a result, we recover the multi-channel Kondo model but with  Luttinger parameters in the leads that are doubled with respect to their bare values. For non-interacting leads, the resulting multi-channel Kondo model has effectively strong interacting reservoirs, with the renormalized Luttinger parameter $K=2$, and the strong coupling QBM fixed point becomes stable in contrast with the standard (non-interacting) Kondo fixed point. The standard multi-channel Kondo model is exactly recovered at the bare value $K=1/2$, {\it i.e.} for strong electronic repulsion in the leads. Interestingly, we also show that for a large number of channels, the model is characterized by two stable fixed points in a certain window of Luttinger parameters, at strong and finite coupling, separated by an unstable fixed point, and thereby predicting a quantum phase transition as the electron tunneling amplitude to the lead is varied. The finite coupling fixed point is in fact analytically connected to the infrared fixed point of the multi-channel Kondo model, and we thus expect similarily that it is not robust against channel asymmetries or against lifting the charge degeneracy. Most of our perturbative results are based on the QBM analysis of Kane and Yi~\cite{Yi-1998, Yi-2002} which we revisit by focusing on the pseudospin wavefunction. The same model at arbitrary charge degeneracy has been also investigated in a very recent work~\cite{Michaeli2016} where the resonant peak in the conductance was shown to be strongly enhanced at degeneracy, and the strong coupling point was argued to be robust. We reproduce their main results, except for the fact that the renormalization group (RG) analysis in Ref.~\onlinecite{Michaeli2016} is limited to lowest order so that the coexistence of stable fixed points for a number of channels above four is not discussed.

The paper is organized as follows. In Sec.~\ref{sec:model}, we introduce the problem and its bosonization description. We discuss the QBM picture in Sec.~\ref{sec:qbm} and review its application to the topological Kondo model. The charge degenerate point is investigated in Sec.~\ref{sec:degenerate} where a mapping onto the M-CKM is formulated and in Sec.~\ref{sec:phase} where the evolution of the different fixed points is determined. Sec.~\ref{sec:conclusion} concludes.

\section{Model and bosonization}\label{sec:model}
\subsection{Model}
\begin{figure}
\centering{\resizebox{0.6\columnwidth}{!}{
\LARGE{
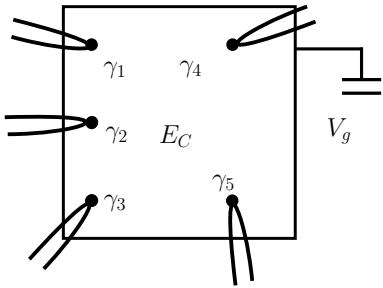}}}
\caption{\label{fig:SchemaBox} Sketch of the Majorana island. A superconducting box is connected through localized Majorana modes $\gamma_{j}$ to $M=5$ normal leads. The Majorana modes are typically realized as boundary bound states of topological nanowires deposited on the superconducting box, and therefore come in pairs. Only the Majorana modes coupled to electronic reservoirs are pictured here.}\end{figure}

We consider the device introduced in Refs. \onlinecite{Beri-2012, Atland-2013} and depicted in Figure~\ref{fig:SchemaBox} composed of a floating mesoscopic superconductor onto which several topological superconducting nanowires have been deposited. Driven in its topologically non-trivial state, each nanowire hosts a pair of zero-energy Majorana bound states located at its extremities. The superconducting island, also called topological Kondo box, is tunnel-coupled via their Majorana bound states  to $M$ normal leads of spinless conduction electrons.

The Hamiltonian describing this device is given by $H=H_\text{box} + H_\text{leads}+H_t$. We focus on low energies, well below the proximity gap induced by the superconducting island on the nanowires, and keep only the state manifold generated by the Majorana operators. The Hamiltonian for the  box thus is simply given by its charging energy  
\begin{equation}
H_\text{box} = E_{C} (\hat{N} - n_g)^2
\end{equation}
with the renormalized backgate voltage $n_{g}$. Formally, $n_g$ is the number of holes on the gate. $E_{C}$ is the charging energy of the box. The number of charges on the box $\hat{N}$, is written as a sum $$\hat{N} = 2 \hat{N}_{c} + \hat{n},$$ where $\hat{N}_{c}$ counts the number of Cooper pairs and $\hat{n}$ the number of fermions in the zero-energy Majorana manifold on the island. $\hat{N}_{c}$ is conjugate to the superconducting phase $\chi$ as expressed by the commutation relation $[ \chi,2 \hat{N}_{c}] =i$. Hence $e^{-2 i \chi}$ is an operator shifting the number of Cooper pairs on the island by $-1$, {\it i.e.} it annihilates a Cooper pair. The Majorana operators satisfy the standard Clifford algebra
\begin{equation}
\gamma_j = \gamma_j^\dagger, \qquad \{ \gamma_j,\gamma_k \} = 2\delta_{jk}
\end{equation}
and can be paired to define fermionic operators, $d_j = (\gamma_{2 j-1} + i \gamma_{2 j})/2$. In this fermionic basis, the Majorana occupation number is simply given by $\hat{n} = \sum_{j=1}^{N/2} d_j^\dagger d_j$, where $N$ (even) is the total number of Majorana bound states on island (we consider $N>M$, see Appendix \ref{App:Tunneling}). For convenience in the notations but without loss of generality, we assume that the first $j=1,\ldots,M$ Majorana are tunnel-coupled to the leads. We are mainly interested in $M\ge 3$, where the system is known to present non-trivial Kondo properties.

The electrons in the mesoscopic box have been polarized due to Zeeman effect\cite{A.Das2012, wire1, wire2, Albrecht2016}. We then assume that the incoming electrons in the leads can penetrate in the box only if they have the right spin polarization. This justifies the representation as semi-infinite one-dimensional spinless fermions of the electrons in each lead. At low energy, the electron field operator in the lead $j$ is $\psi_{j} (x) = e^{i k_F x} \psi^\dagger_{R, j} (x) +  e^{-i k_F x} \psi^\dagger_{L, j} (x)$, introducing right- and left-movers, where $k_F$ is the Fermi momentum. The Hamiltonian has the form
\begin{multline}\label{eq:leadh}
H_\text{leads}=-i v_F \sum\limits_{j=1}^{M} \int\limits_0^{+\infty} dx \left( \psi^\dagger_{R, j} \partial_x \psi_{R, j}\right. \\
-\left.\psi^\dagger_{L, j} \partial_x \psi_{L, j} \right) + H_{\rm int}
\end{multline}
where $v_F$ is the Fermi velocity. $H_{\rm int}$ contains electron-electron intra-wire interaction and will be included as a Luttinger parameter $K$ in the bosonization procedure\cite{Haldane1981, QP1DGiamarchi2004, BosonizationGogolin1998}.
Finally, the coupling between the Majorana bound states and the leads are described by the tunneling Hamiltonian (see Appendix \ref{App:Tunneling} for a proper derivation)
\begin{equation}\label{eq:tunnel}
H_t=-\sum\limits_{j=1}^{M} t_{j}  e^{-i \chi} \psi^\dagger_{j}(0)\gamma_{j} +h.c.
\end{equation}
$t_{j}$ are the tunneling amplitudes, all taken to be real and positive. The symbol $(0)$ refers to the position $x=0$ of each wire coupled to the island.

\
\subsection{Bosonization and Majorana fermions}\label{subsec:boson}
We take advantage of the one-dimensional character of lead electrons and apply Abelian bosonization\cite{Haldane1981, QP1DGiamarchi2004, BosonizationGogolin1998}. The Klein factors introduced in the bosonization procedure can be combined with the impurity Majorana fermions to derive a purely bosonic Hamiltonian. The ensuing technical analysis is considerably simplified.\\

Introducing a short-distance length $\alpha$, the Abelian bosonization\cite{QP1DGiamarchi2004} formula expresses the fermion field operator (for semi-infinite leads)
\begin{equation}
\psi_{R/L,j}=\frac{U_{R/L,j}}{\sqrt{2\pi \alpha}} e^{-i(\pm\phi_{j}-\theta_{j})},
\end{equation}
in terms of two conjugate boson fields 
\begin{equation}
[\phi_{j}(x), \theta_{k}(x')]=i\frac{\pi}{2}\text{sgn}(x'-x) \delta_{j,k},
\end{equation}
and Klein factors satisfying 
\begin{equation}
\{U_\gamma, U_{\gamma'}^\dagger\}= 2\delta_{\gamma,\gamma'}, \qquad \gamma = R/L,j.
\end{equation}
In the thermodynamic limit, Klein factors can be identified with Majorana fermions $U_\gamma \simeq U_\gamma^\dagger$.
In this representation, the lead Hamiltonian~\eqref{eq:leadh} is written as
\begin{align}
H_\text{leads}&= \sum\limits_{j} \int_0^{+\infty} dx \frac{v_{F}}{2\pi}(K(\partial_x \theta_{j})^2 + \frac{1}{K}(\partial_x \phi_{j})^2) \label{Eq:leadhBoson}.\\
&=\sum\limits_{j} H_0\{\phi_j, \theta_j, K\} 
\end{align}
The Luttinger parameter accounts for the electron-electron interaction in the leads. If they are repulsive, $K<1$. The leads stop at $x=0$ in the vicinity of the island, imposing the Dirichlet boundary condition $\psi_{R,j} (0) = \psi_{L,j} (0)$. Hence there is a single Klein factor per lead, $U_{R,j} = U_{L,j}$, and $\phi_{j} (0) = 0$. This can also be understood the following way: instead of considering a wire only from $-\infty$ to $0$, we separate the left- and right-moving modes and consider only a single chiral fermionic wire going from $-\infty$ to $\infty$, with the box/impurity at $x=0$. \\

\section{Quantum Brownian Motion and Topological Kondo model}\label{sec:qbm}

\subsection{Quantum Brownian Motion}

The quantum impurity model we derived in the previous section is a boundary one-dimensional model. By integrating all modes with $x>0$ except the operator at $x=0$ in the action formalism, it can be formulated as a zero-dimensional spatial problem with one temporal dimension. The integration of these Gaussian bosonic modes\cite{QP1DGiamarchi2004} transforms the Euclidean action into
\begin{equation}
S =  \sum_{\omega_m} \sum_{j} \frac{|\omega_m| K }{2 \pi \beta} |\theta_{j} (\omega_m)|^2 + \int_0^\beta d \tau H_{\rm P} (\tau)
\end{equation}
where $\beta$ is the inverse temperature and all bosonic fields $\theta_{j} (\omega_m)$ are implicitly taken at $x=0$, and
\begin{equation}
H_{\rm P} = H_{\rm box} + H_t. 
\end{equation}
We have introduced the bosonic Matsubara frequencies $\omega_m$ over which the  action is summed. The first term in this expression $\sim |\omega_m|$ describes dissipation caused by electron-hole excitations in the leads. As we will show in the following, the global mode $(\Phi, \Theta)$,  $\Theta/\Phi=\frac{1}{\sqrt{M}} \sum_{j} \theta_{j}/\phi_j$, separates from the other modes. In a general fashion, we introduce the $M-1$ dimensional bosonic modes $\vec{r}$ and $k$ defined by $(\vec{r}, \Theta)=R \vec{\theta}$, $(\vec{k}, \Phi)=R \vec{\phi}$, $R$ being an orthogonal matrix (see an example in Appendix \ref{App:Rmatrix}). Dropping the global mode, the action can be rewritten as
\begin{equation}\label{eq:action}
S =  \sum_{\omega_m} \sum_{j}^{M-1} \frac{|\omega_m| K}{2 \pi \beta }|r_{j} (\omega_m)|^2 + \int_0^\beta d \tau H_{\rm P}  (\tau)
\end{equation}
The action can be identified with the QBM model\cite{Yi-1998, Yi-2002} where a massless particle  subject to dissipation moves in a $M-1$ dimensional space with coordinates $\vec{r}$.  The potential $H_{\rm P}$ seen by the particle depends not only on the coordinate  $\vec{r}$  but also on the charge configuration $\hat{N}$. Depending on the gate voltage, we shall restrict the charge to a single value, in which case we have a scalar potential, or two degenerate values represented by a ficticious spin attached to the particle. Following the seminal approaches of Refs.~\onlinecite{Yi-1998, Yi-2002} with the dual action of instanton tunneling~\cite{schmid1983,Kane1992}, we shall use this analogy to describe the low energy properties of the model in the strong coupling limit.

\subsection{Far from charge degeneracy: the Topological Kondo limit}\label{sec:TKL}

Before discussing the degenerate case, let us shortly review the topological Kondo model~\cite{Eriksson-2014, Eriksson-2014-Manifold, Beri-2012, Beri-2013, Galpin-2014, Plugge-2016,  Zazunov-2014, Atland-2013, Atland-2014} and its QBM solution. This will introduce concepts and notations that will be useful in the analysis of the degenerate case.

We begin by assuming that the charging energy is the dominant energy scale $E_C \gg T, t_j^2/v_F$ such that only one or two charge states are relevant for transport. We further assume that $n_g$ is close to an integer value $n$ and project the model onto the charge quantized configuration $\hat{N} = n$. Given that $H_t$ changes the number of electrons on the island by $\pm 1$, the Schrieffer-Wolff expansion allows us to take into account virtual processes through the neighbouring charge states $n+1$ and $n-1$. To second order, we obtain the exchange term
\begin{equation}\label{eq:sw}
H_{SW}=\sum\limits_{j\neq k}^{M, M}  \tilde{\lambda}_{j,k}  \psi^\dagger_{k} (0) \psi_{j}  (0) \gamma_{j} \gamma_{k},
\end{equation}
with $\tilde{\lambda}_{j,k}=\tilde{t}_j \tilde{t}_k \left(\frac{1}{\Delta E(n+1)}-\frac{1}{\Delta E(n-1)} \right)$ and 
\begin{equation}
\Delta E(n') = E_C (n'-n)(n+n'-2 n_g)
\end{equation}
the difference in energy $H_{\rm box}$ between the charge value $n'$ and $n$.  The Schrieffer-Wolff transformation also produces small scattering potential terms $\sim \psi^\dagger_{j} \psi_{j}$ that do not change under renormalization group and can be discarded. We note that the tunneling amplitudes in the Schrieffer-Wolff Hamiltonian~\eqref{eq:sw} are in fact renormalized~\cite{Zazunov2011} by the RG process between the short-time cutoff $\tau_c = \alpha/v_F$ and the charging energy $E_C$ where they increase with the scaling exponent $1-1/(2 K)$ such that $\tilde{t}_j = \sqrt{E_C/\tau_c} \, t_j (\tau_c/E_C)^{1-1/(2 K)}$.

In terms of our bosonized fields, the exchange term is reexpressed as
\begin{equation}\label{eq:sw1box}
H_{SW}=\sum\limits_{j\neq k}^{M, M}  \lambda_{j,k}  U_k U_j \gamma_{j} \gamma_{k} e^{i(\theta_j-\theta_k)},
\end{equation}
with the notation $\theta_{j} \equiv \theta_{j} (0)$ and $\lambda_{j,k}=\frac{\tilde{\lambda}_{j,k}}{\pi\alpha}$. The product $p_{j} =  i \gamma_{j} U_{j}$ is the parity operator associated with the Majorana fermions $\gamma_{j}$ and $U_{j}$. The different $p_{j}$ commute between themselves and with the Hamiltonian, have eigenvalues $\pm 1$, but do not conserve the full parity operator. However, the $M-1$ independent products $p_j p_k = \pm 1$ can be diagonalized simultaneously. The sign of $p_{j} p_{k}$ can be changed by shifting the bosonic fields by $\pi$, we thus fix it arbitrarily to $+1$. With these conventions and choice of gauge, the exchange term assumes a fully bosonic form
\begin{equation}
H_{SW}=-\sum\limits_{j \neq k}^{M,M} \lambda_{j,k} \cos(\theta_j-\theta_k),
\end{equation}
The global mode $(\Phi,\Theta)$, $\Phi/\Theta=\frac{1}{\sqrt{M}} \sum \phi_j/\theta_j$ decouples from $H_{SW}$ as announced.

One can compute the poor man's scaling equations for this problem\cite{Anderson1970-poorman}, and the renormalization group (RG) analysis is straightforward
\begin{equation}
\frac{d \lambda_{j,k}}{d \ell} = \left(1-\frac{1}{K} \right) \lambda_{j,k} + 2 \sum\limits_{m\neq j,k}^M \lambda_{j,m} \lambda_{m,k}, \label{eq:RGTK}
\end{equation}
with the flow parameter $\ell = \ln \tau_c$. Channel asymmetry between the different electron hopping terms $\lambda_{j,k}$ is not relevant and the RG flow points to a symmetric combination $\lambda_{j,k} \to \lambda (1-\delta_{j,k})$. Assuming channel symmetry reduces the RG equation to
\begin{equation}
\frac{d \lambda}{d \ell} = \left(1-\frac{1}{K} \right) \lambda + 2 (M-2) \lambda^2, \label{eq:RGTK2}
\end{equation}
where three fixed points can be identified. First, the weak coupling fixed point with $\lambda=0$, noted $(O)$, corresponding to decoupled leads between which no electric current flows, or, using the Kubo formula detailed in  Appendix \ref{App:Kubo}, 
\begin{equation}\label{eq:kubo}
G_{j,k}^{(O)} = 0,
\end{equation} 
where  $G_{j,k}$ is the conductance between the leads $j$ and $k$. This is an attractive point for $K< 1$. 

For $K\geq1$, the growth of $\lambda$ under renormalization suggests to study a strong coupling limit $\lambda = +\infty$ noted $(S)$. In this limit, the fields $\theta_j$ are pinned to one of the minima of the potential described by $H_{SW}$. Using the Kubo formula, the conductance\cite{Yi-1998, Yi-2002} $G_{j,k}$ is now given by (Appendix \ref{App:Kubo} for a derivation)
\begin{equation}
G_{j,k}^{(S)} = \frac{2 e^2 K}{h}\left( \frac{1}{M}-\delta_{j,k} \right). \label{eq:condtopo}
\end{equation}
In agreement with the physical picture of strong coupling, this is the maximum conductance one can reach with the constraint of charge conservation. It indeed corresponds to a perfect symmetric transmission of incoming electrons. For $M=2$, we recover that the Majorana Kondo box maps onto a problem of resonant tunneling, where the conductance is simply $\frac{ e^2 K}{h}$ for spinless fermions. The factor $\left( \frac{1}{M}-\delta_{j,k} \right)$ can be understood in the following way: due to the isometry of the fixed point, an electron arriving on the impurity is scattered uniformly in all leads, imposing
\begin{equation}
G_{j,k}^{(S)} = G-\delta_{j,k} G_0, 
\end{equation}
where $G_0 = 2 e^2 K/h$.
Conservation of the current leads to $\sum_j G_{j,k}=0$, that is to say $G_0/G=M$, giving us the aforementioned factor. We note that in practice the Luttinger liquid wires are ultimately contacted to Fermi liquid reservoirs which has the effect of renormalizing~\cite{Safi1995, Maslov1995} the Luttinger parameter to $K=1$ in Eq.~\eqref{eq:condtopo}.

A third fixed point, noted (I), is identified for $K<1$ corresponding to the intermediate coupling
\begin{equation}
\lambda^I = \frac{1/K-1}{2(M-2)}.
\end{equation}
It is unstable against both weak (O) and strong coupling (S) fixed point. The perturbative RG equation~\eqref{eq:RGTK2} justifies the existence of this intermediate unstable point only for $K$ close but below $1$, such that $\lambda^I$ remains small.

To check the stability of $(S)$, one can perform an instanton analysis. Given the simple structure of the potential, the more relevant/less irrelevant operators at $(S)$ are operators translating one minimum of the potential described by $H_{SW}$ to one of its neighbours. The variable $\Theta$ does not appear in $H_{SW}$ and therefore has a free evolution reflecting charge quantization~\cite{Zazunov-2014} on the island. For clarity, we henceforth set $\Theta=0$. In terms of the variable $\vec{r}$, the minima of the potential described by $H_{SW}$ form a (hyper)triangular lattice, and we can explicitly construct the operators connecting them. For simplicity, we write these minima in the $\theta_j$ basis. A minimum, $R_0$, and all its nearest-neighbours $R_k$, $(k=1...M)$ are given by
\begin{equation}\label{eq:config}
\begin{split}
R_0&:\quad \theta_j=0\quad \forall j \\
R_k&:\quad \theta_k=\frac{2 \pi (M-1)}{M}, \quad\theta_j=-\frac{2 \pi }{M}\quad \forall j \neq k \\
\end{split}
\end{equation}
In a semi-classical analysis, quantum fluctuations around these minima are neglected and the only low energy processes are instanton solutions connecting them. Introducing the variable $\phi_k$ (the charge in lead $k$) canonically conjugated to $\theta_k/\pi$,
\begin{equation}
[\phi_j,\theta_k] = i \pi \delta_{j,k},
\end{equation}
it is then possible to explicitely construct the instanton operators. The shift from $R_0$ to $R_k$ is thus realized by the translation operator
\begin{equation}\label{eq:ok}
\hat{O}_k=\exp \left[ 2i \left(\phi_k-\frac{1}{\sqrt{M}} \Phi \right) \right]
\end{equation}
where $\Phi = \frac{1}{\sqrt{M}} \sum_k \phi_k$ is the total charge field. Identifying $\phi_k$ with the field $\phi_k (0)$, we obtain the following dual action describing the vicinity of $(S)$
\begin{equation}\label{eq:strong}
S =  \sum_{\omega_m} \sum_{j}  \frac{|\omega_m| }{2 \pi K \beta} |\phi_{j} (\omega_m)|^2 
 - v \int_0^\beta d \tau \sum_k \hat{O}_k (\tau), 
\end{equation}
where $v$ describes weak backscattering of electrons coming from reservoirs.

The stability of $(S)$ is controlled by the dimension of the operators $\hat{O}_k$. Using the free part of the action~\eqref{eq:strong}, one obtains the dimension $2 K (M-1)/M$. This result can also be understood in the QBM picture~\cite{Yi-1998, Yi-2002} where the  product between the dimension of the original perturbation $e^{i (\theta_j-\theta_k)}$ (here $1/K$) and the leading irrelevant operator at strong coupling is fixed to $2 (M-1)/M$ for a hypertriangular lattice in $M-1$ dimension. We therefore find that $(S)$ is stable for $K> \frac{M}{2(M-1)}$, and unstable towards $(O)$ for smaller values of $K$. Computing the one-loop RG equation, one obtains~\cite{Yi-2002} that the unstable fixed point $(I)$ departs from $(S)$ for $K> \frac{M}{2(M-1)}$ and disappears below. 

\begin{figure}
\centering{\includegraphics[width=\columnwidth]{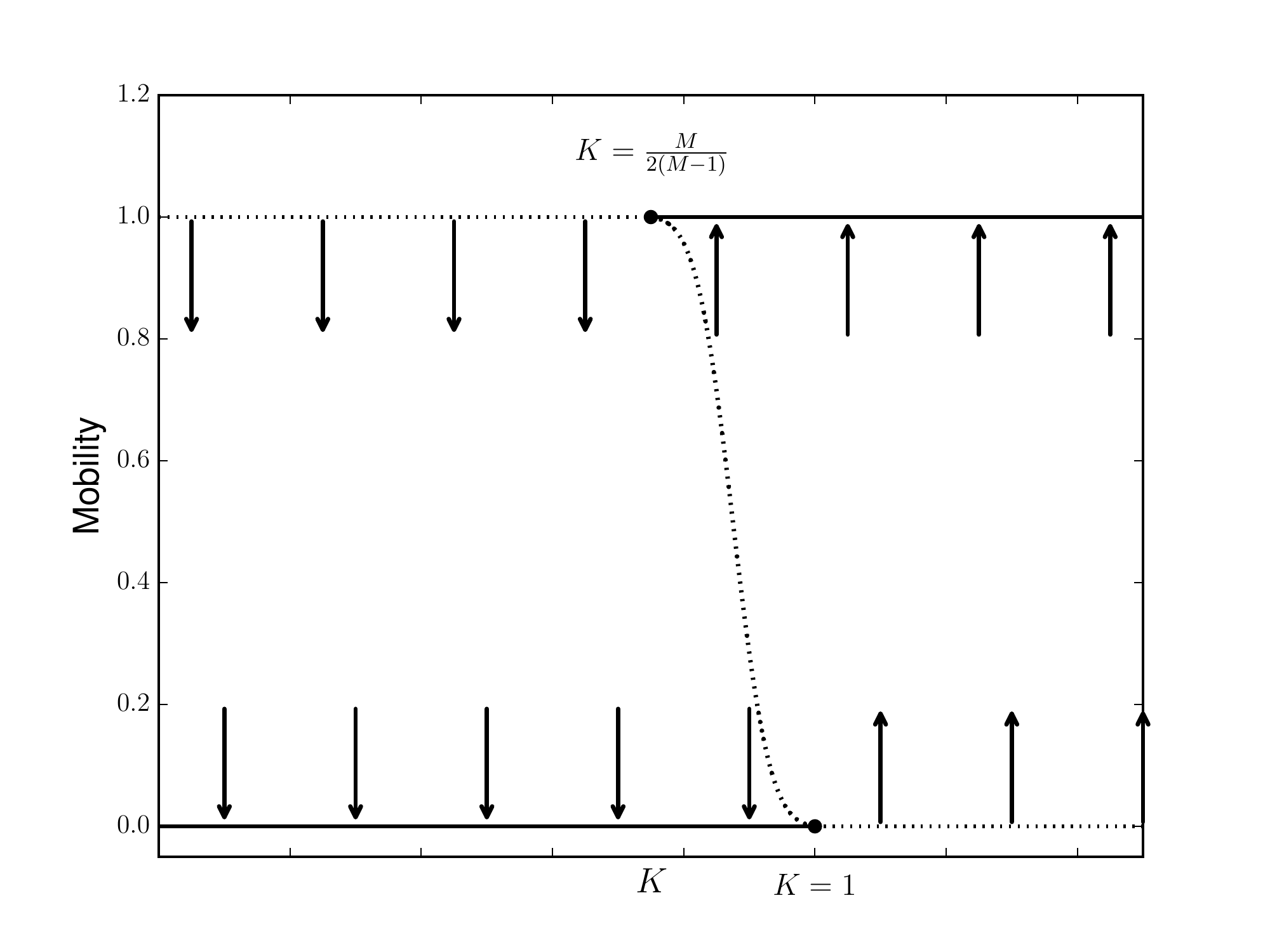}}
\caption{Phase diagram and flow for the single Majorana Kondo box far from charge degeneracy. 
The mobility is defined by: $\mu=\frac{hG_{j,j}}{2e^2} \frac{M}{1-
M}$. Solid line describes stable fixed points while pointed lines are unstable. The arrows depict the flow of the most relevant/less irrelevant operators of the corresponding fixed point.}
\label{fig:PD1_Box}
\end{figure}

The whole phase diagram for the topological Kondo model is summarized in Fig.~\ref{fig:PD1_Box}. For $\frac{M}{2(M-1)}<K<1$, a first order transition between zero and maximum conductance is predicted~\cite{Beri-2012} to occur as the coupling to the reservoir is varied. The Kondo temperature is evaluated from the RG equation~\eqref{eq:RGTK2},
\begin{equation}
T_K \simeq  E_c e^{-\frac{1}{2(M-2) \nu \lambda}},
\end{equation}
 where $\nu$ is the density of states in the wires, sets the crossover energy scale between weak and strong coupling.

\section{Charge degeneracy point: an exact mapping to the multichannel Kondo Model}\label{sec:degenerate}

We now turn to the charge degenerate case where the gate voltage is fixed to a half-integer value $n_g = n+1/2$. The two charge states $n$ and $n+1$ are energetically equivalent and define a low energy sector  akin to a spin-$1/2$ Hilbert space. Further assuming a large charging energy $E_C \gg T, t_j^2/v_F$, we project to this subspace and rewrite the full bosonized Hamiltonian as
\begin{equation}\label{eq:onecharge}
H=H_{\rm leads} -  \sum\limits_{j=1}^{M} \frac{2 t_j}{\sqrt{2 \pi \alpha}}\left(  \tau_- U_j \gamma_{j,r} e^{-i \theta_j} +h.c. \right),
\end{equation}
with the pseudo-spin operator $\tau_- |n+1\rangle = |n\rangle$. Similarly to Eq. \ref{eq:sw}, the hopping terms are renormalized when reducing the charge sector to an effective pseudo-spin. In terms of the original bare values of the tunneling, they can be reexpressed as: $\tilde{t}_j \approx t_j \left(\tau_c /E_c \right)^{1-\frac{1}{2K}}$. In the following, for simplicity, we drop the tilde on $t_j$.

We show in the following that this Hamiltonian can be exactly mapped onto the M-CKM. First, one can rescale the bosonic fields in order to obtain the correct dimension for the operators coupled to the pseudo-spin, namely
\begin{equation}\label{eq:equiv}
\tilde{K}=2K,~~~\tilde{\theta}_{j} = \frac{\theta_j}{\sqrt{2}},~~~\tilde{\phi}_j=\sqrt{2} \phi_j
\end{equation}
Second, we use the trick presented in the previous section to fuse the Majorana fermions and the Klein factors. We introduce the operators $p_j = i \gamma_j U_j$ and fix them to $1$. Shifting the $\theta_j$ variables by $\pi/2$ to absorb an $i$ factor, we obtain the bosonized form
\begin{multline}\label{eq:boso}
H=H_\text{leads}\{\tilde{\phi}, \tilde{\theta}, \tilde{K}\} - \sum\limits_{j=1}^{M} \left( \frac{J_{\perp, j}}{2} \tau_+ e^{i \sqrt{2} \tilde{\theta}_j} +h.c \right)\\
 + \frac{v_F}{\sqrt{2}} J_z \tau_z \sum\limits_{j=1}^{M} \partial_x \tilde{\phi}_j,
\end{multline}
where $J_{\perp,j}= \frac{4 t_j}{\sqrt{2 \pi \alpha}}$ and $J_z=0$, corresponding to the spin sector of the anisotropic M-CKM (the charge mode decouples from the impurity spin). Alternatively, contact with the M-CKM can be made from Eq.~\eqref{eq:onecharge} with the analogy 
\begin{equation}
U_{j, \uparrow}= U_j,~~~~ U_{j, \downarrow}= \gamma_j.
\end{equation}
We note that the Luttinger parameter $\tilde{K}$ in Eq.~\eqref{eq:boso} characterizes the spin sector and requires in the M-CKM the SU(2) spin symmetry to be broken in the leads to be different from one. Here, no such symmetry-breaking is necessary as Eq.~\eqref{eq:boso} emerges as an effective model with $\tilde{K} \ne 1$ as a general case. For non-interacting leads for example, where  $K=1$, one has $\tilde{K} = 2$.

The effective Kondo model~\eqref{eq:boso} is strongly anisotropic since $J_z=0$. A finite $J_z$ is nevertheless generated in the RG process. We consider for simplicity the channel-isotropic case $J_{\perp,j}=J_\perp$, and derive the corresponding RG equations following Anderson, Yuval and Hamman~\cite{Anderson1970} (see Appendix \ref{App:RG-MCKM})  extended to the interacting case $\tilde{K} \ne 1$,
\begin{align}
\frac{d J_z}{d \ell}&=J_\perp^2 \left(\frac{1}{\tilde{K}}-\frac{M}{2} J_z\right), \label{eq:rg3} \\
\frac{d J_\perp}{d \ell} &= \left(1-\frac{1}{\tilde{K}} \right) J_\perp + J_z J_\perp \left(1-\frac{M \tilde{K}}{4} J_z\right).\label{eq:rg4}
\end{align}
These equations are perturbative in $J_\perp$ and exact in $J_z$. Studying these equations, one sees that the longitudinal coupling $J_z$ is attracted by the fixed point value $J_z = \frac{2}{M \tilde{K}}$ at which $J_z$ ceases to be generated and the RG evolution of $J_\perp$ decouples. It corresponds in fact to the standard Emery-Kivelson\cite{Emery1992} (or Toulouse\cite{Toulouse1970}) limit in the M-CKM~\cite{Matveev1991,Yi-1998,Yi-2002}. It is reached by the RG flow even if the initial value of $J_z$ is zero. Therefore, it makes sense to start with the model~\eqref{eq:boso} directly at $ J_z = \frac{2}{M \tilde{K}}$ and perform the unitary transformation
\begin{equation}\label{eq:unitary}
U = \exp \left( i \frac{\tilde{K} J_z \sqrt{M}}{\sqrt{2}} \tilde{\Theta} (0) \tau_z \right),
\end{equation}
to eliminate the $J_z$ term from Eq.~\eqref{eq:boso}. The resulting Hamiltonian is 
\begin{multline}\label{eq:kondo}
\hat{U}^\dagger H \hat{U}=H_\text{leads}\{\phi, \theta, K \} \\
- \sum\limits_{j=1}^{M} \left( \frac{J_{\perp, j}}{2} \tau_+  e^{i (\theta_j-\frac{1}{\sqrt{M}} \Theta) } +h.c \right)
\end{multline}
when written again in terms of the old fields, see Eq.~\eqref{eq:equiv}. The use of the unitary transformation is not only for mathematical convenience but also possesses a physical significance. The model~\eqref{eq:kondo} is now invariant under a global shift of all fields $\theta_j$ which implies that the total mode $\Theta = \frac{1}{\sqrt{M}} \sum_{j} \theta_{j}$ decouples as in the previous section. Although the present degenerate case does not satisfy a strict charge quantization, the flow of incoming electrons must exactly compensate the flow of outgoing electrons since there can be no charge accumulation in the floating quantum box. As a result, current conservation also holds as reflected by the free mode $\Theta$. 

\section{Phase diagram at charge degeneracy}\label{sec:phase}

The pertubation operators in Eq.~\eqref{eq:kondo} have dimension $\frac{M-1}{2 M K}$ after the unitary transformation. Hence, For $K<\frac{M-1}{2M}$, the system flows towards the uncoupled fixed point $(O)$, with $J_\perp=0$, and the conductance is zero as in Eq.~\eqref{eq:kubo}. For $K > \frac{M-1}{2M}$, $(O)$ is unstable and the RG equations~\eqref{eq:rg3},~\eqref{eq:rg4} must be supplemented by the next order in $J_\perp$. Eq.~\eqref{eq:rg3} is unchanged whereas Eq.~\eqref{eq:rg4} becomes
\begin{equation}\label{eq:RGMCKM}
\begin{split}
\frac{d J_\perp}{d \ell} = & \left(1-\frac{1}{2 K} \right) J_\perp \\
& + J_z J_\perp \left(1-\frac{M K}{2} J_z\right) - C_M (K) J_\perp^3,
\end{split}
\end{equation}
where $C_M (K) = {\cal O}(1)$ depends on $K$ and the number of channels $M$. The coefficient $C_M^* = C_M [(M-1)/(2 M)]$ can be evaluated at the threshold for the instability of the uncoupled point $(O)$ and shown to be always positive~\cite{Yi-2002}, demonstrating an intermediate RG-stable fixed point at $J_z^{I} = 1/(M K)$ and 
\begin{equation}
J_\perp^I = \sqrt{ \frac{K-\frac{M-1}{2 M }}{K C_M^*}},
\end{equation}
valid for $K$ close to $\frac{M-1}{2M}$. At the fixed point $(I)$, the conductance is non-vanishing, $\propto (J_\perp^I)^2$ for small $J_\perp^I$. It increases continuously with the Luttinger parameter $K$. The way $(I)$ connects with the strong coupling fixed point $(O)$ depends on the value of $M$ and shall be discussed below where the strong coupling limit is investigated.  At the specific point where $K=1/2$, then $\tilde{K} = 1$ and Eq.~\eqref{eq:kondo} represents exactly the non-interacting M-CKM. From conformal theory\cite{Affleck1993, Affleck1995, Affleck-2001}, it is known that the conductance is given in that case by 
\begin{equation}
 G_{j,k}=\frac{2e^2 K \sin^2(\frac{\pi}{M+2})}{h}\left(\frac{1}{M} -\delta_{j,k}\right).
 \end{equation} 
The conductance for other values of $K$ is not known analytically.

Let us study the model at strong coupling. In Ref.~\onlinecite{Yi-2002}, Eq.~\eqref{eq:kondo} was argued to be the dual model of a particle moving in a hyperhoneycomb lattice formed by two interpenetrating triangular lattices between which the operators $\tau_{\pm}$ alternate. We discuss here directly the strong coupling limit, noted $(0)$, $J_{\perp, j} = +\infty$ of Eq.~\eqref{eq:kondo} and construct explicitely its dual action by taking into account the pseudo-spin wavefunction. In the spirit of a semi-classical approach, we minimize Eq.~\eqref{eq:kondo} (without the lead term) with respect to the fields $\theta_j$ and the spin configuration, whereas the total field $\Theta$ factorizes (set to zero for simplicity) and is free. In the channel-isotropic case, $J_{\perp, j}= J_\perp$, the energy to minimize has the form
\begin{equation}
- J_\perp \begin{pmatrix} 0 & {\cal S} \\ {\cal S}^* & 0 
\end{pmatrix} \qquad {\cal S} = \sum_j e^{i (\theta_j-\frac{1}{\sqrt{M}} \Theta) } 
\end{equation}
Interestingly, the minima are located  at exactly the same field $\theta_j$ positions as in the topological Kondo model, $R_0$ and its neighbours $R_k$ as given in Eq.~\eqref{eq:config}, forming a triangular lattice in a $M-1$ dimensional space orthogonal to the total mode $(1,1,\ldots,1)$ direction. But the problem is nevertheless different since there is an additional pseudo-spin degree of freedom, and each minimum is characterized by a certain spin wavefunction,  $(|+\rangle + |-\rangle)$ for $R_0$, and $(e^{- i \pi/M} |+\rangle + e^{i \pi/M} |-\rangle)$ for $R_k$. The conductance at strong coupling is still given by Eq.~\eqref{eq:condtopo}.

\begin{figure}
\begin{center}
\subfigure{\includegraphics[width=\columnwidth]{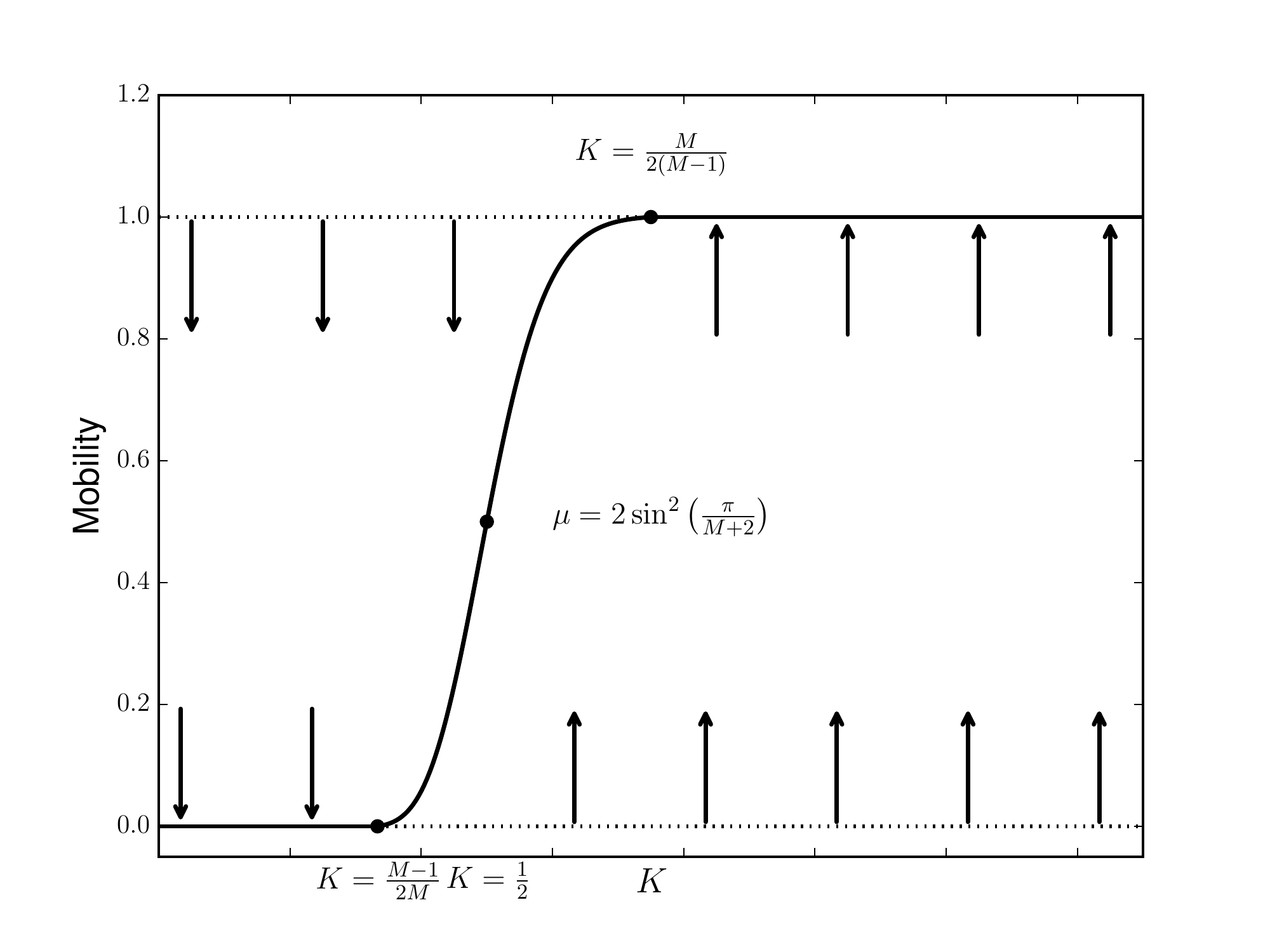}}
\subfigure{\includegraphics[width=\columnwidth]{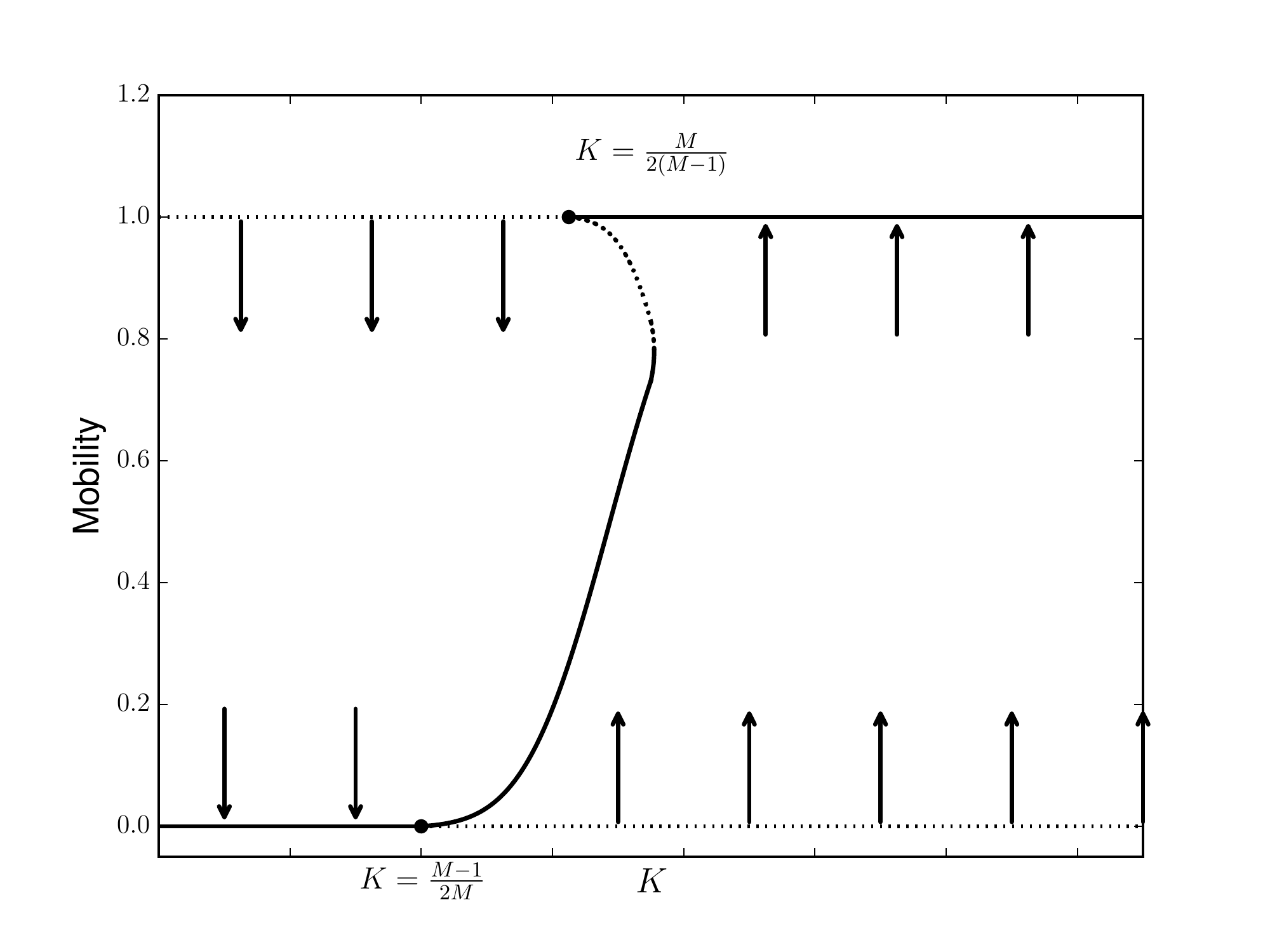}}
\end{center}
\caption{Phase diagrams and flows for the single Majorana Kondo box at charge degeneracy. The mobility is defined by: $\mu=\frac{hG_{j,j}}{2e^2} \frac{M}{1-
M}$. The first graph describes the simple case of $M=3,4$ while the second is a schematic of what happens for $M\ge5$.}
\label{fig:PD2_Box}
\end{figure}

Technically, moving from one minimum to its neighbour rotates the spin direction by an angle $2 \pi/M$ around the $z$ axis. Hence, performing a loop starting and ending at $R_0$ exhausting the different neighbour directions, one obtains a rotation of $2 \pi$ coming with an overall phase $e^{i \pi}=-1$ resulting from the pseudo-spin Berry phase~\footnote{The $-1$ factor was also interpreted as an effective and alternating $\pm \pi$ flux threading each plaquette of the triangular lattice~\cite{Yi-2002}.}.  This sign is in fact responsible for the change of sign of the second order term in the RG flow for $M=3$ (and the third order term for $M=4$), leading to differing phase diagrams for the triangular and honeycomb lattices. The dual action, representing the instanton solutions connecting the minima of Eq.~\eqref{eq:kondo}, is constructed in the same way as in Sec.~\ref{sec:TKL}. The leading irrelevant operators at low energy are thus given by the translations
\begin{equation}\label{eq:trans}
\hat{O}_k^{(h)} = \exp \left[ 2i\left(\phi_k-\frac{1}{\sqrt{M}} \Phi\right)\right] \exp \left(-\frac{i \pi}{M} \tau_z \right),
\end{equation}
where the second part accounts for the spin rotation between two consecutives minima, {\it e.g.} $R_0$ and $R_k$. Its dimension is the same as $\hat{O}_k$ in Eq.~\eqref{eq:ok}, $2 K (M-1)/M$. We note that the minima of the potential $H_{\rm dual} = - v \sum_j \hat{O}_k^{(h)}$ form a hyperhoneycomb lattice for a given spin projection $\tau_z = +1$ or $-1$. The RG analysis of the model at strong coupling, or $v \ll 1$, depends on the dimension $M$~\cite{Yi-2002}. For $M=3$, the RG equation is
\begin{equation}
\frac{dv}{d\ell} =  (1-4 K/3)v-2v^2
\end{equation}
where the last term sign has its origin in the pseudo-spin Berry phase as discussed in the appendix~\ref{App:RG-MCKM}. As a result, the intermediate coupling fixed point $(I)$ occurs for $K<3/4$, with $v^{(I)} = \frac{1-4 K/3}{2}$ valid for $1-4 K/3 \ll 1$, and is stable. Its is continuously connected to the intermediate fixed point $(I)$ found at weak coupling. For $M=4$, one obtains
\begin{equation}
\frac{dv}{d\ell} =  (1-3 K/2 )v- (A_4 - B_4) v^3, \qquad A_4 > B_4
\end{equation}
leading to $(I)$ for $K<2/3$ and $v^{(I)} = \sqrt{\frac{1-3 K/2}{A_4 - B_4}}$. In both situations, $M=3$ and $4$, the phase diagram has the form shown in the upper panel of Fig.~\ref{fig:PD2_Box} and the stable intermediate point $(I)$ connects the weak and strong coupling fixed points as $K$ is varied.

This is contrast with $M \ge 5$ where the RG equation takes the form 
\begin{equation}\label{eq:RG5}
\frac{dv}{d\ell} =  \left(1-\frac{2(M-1) K}{M} \right)v + B_M v^3, 
\end{equation}
with $B_M >0$. The intermediate fixed point exists for $K > 2 (M-1)/M$ and is unstable. This suggests the phase diagram represented in the lower panel of Fig.~\ref{fig:PD2_Box}. Comparing with the non-degenerate case, or topological Kondo model, one observes that exactly the same RG equation~\eqref{eq:RG5} holds~\cite{Yi-2002}. The first reason is that the fields $\theta_j$ are pinned at the same positions irrespective of charge degeneracy. Moreover, the spin wavefunction, and the corresponding Berry phase, plays a role for a product of at least $M-1$ $\hat{O}_k^{(0)}$ operators such that perturbation theory differs  only for orders above $M-2$. The result is that the critical line $(I)$ is the same at high mobility in both non-degenerate and degenerate cases for sufficiently large $M$, {\it i.e.} the departures of the dotted lines in Figure~\ref{fig:PD1_Box} and Figure~\ref{fig:PD2_Box} (lower panel) from strong coupling (mobility $\mu=1$) are identical. The two curves start to differ at larger $v$ (or smaller mobility) where the line $(I)$ at charge degeneracy is below the topological Kondo case~\footnote{Nothing prevents the curve at charge degeneracy to cross above the topological Kondo line (non-degenerate case) for even lower mobility before turning over.}. At even smaller mobility, the effect of the pseudo-spin Berry phase becomes prominent: the line $(I)$ eventually turns over and connects with the stable fixed-point line $(I)$ originating from weak coupling and containing the multi-channel Kondo fixed point at $K=1/2$, see Figure~\ref{fig:PD2_Box} (lower panel).

The spin wavefunction also provides a physical picture to understand the effect of a small charge degeneracy $\delta \ll 1$, with
\begin{equation}
n_g = n +1/2+ \delta.
\end{equation}
For example, at strong coupling, the semi-classical energy to minimize is
\begin{equation}
- J_\perp \begin{pmatrix} 2 \delta E_C & {\cal S} \\ {\cal S}^* & - 2 \delta E_C
\end{pmatrix},
\end{equation}
with eigenvalues $\pm \sqrt{4 (\delta E_C)^2+|{\cal S}|^2}$. The fields $\theta_j$ are thus pinned at the same positions $R_{0,k}$ but the spin wavefunction is polarized by $\delta \ne 0$ along the $z$ direction, thereby reducing the impact of the Berry phase. Since $\delta$ is relevant on the $(I)$ critical line, as we know from the M-CKM at $K=1/2$, this implies that the system flows at low energy towards the non-degenerate case, or Figure~\ref{fig:PD1_Box}. At finite energy (temperature), we expect a continuous crossover for the $(I)$ critical line between the two limiting cases represented by  Figure~\ref{fig:PD1_Box} and Figure~\ref{fig:PD2_Box}.

Finally, from Eq. \ref{eq:RGMCKM}, one can evaluate the one-loop Kondo temperature:
\begin{equation}
T_K \approx \nu t^{\frac{2K}{2K-1}},
\end{equation}
where $\nu$ is the density of states and $t$ is the typical bare value of the tunneling term. For $\frac{1}{2}<K<K_c$, it corresponds to the Kondo temperature of the MCKM intermediate fixed point, while for $K>K_c$, it characterizes the strong coupling limit. When the leads are non-interacting ($K=1$), we obtain the very physical expression: $T_K \approx \nu t^2$, characteristic of a resonant tunneling transport through the superconducting island, in agreement with Ref.~\onlinecite{Michaeli2016}.

Up to know we only considered flavor isotropic tunneling, \textit{i.e.} $t_j=t~\forall j$. While flavor anisotropy was irrelevant in the Topological Kondo model, it is no longer the case at charge degeneracy\cite{Affleck1992, Zarand2000} where we are dealing with an intermediate fixed point. Indeed, the critical model goes from M-CKM to M'-CKM, where M' is the number of channels with the largest value of $J_{\perp,j}$ (generally $M'=1$). Consequently, observing the fractional non-trivial conductance of M-CKM will require fine-tuning~\cite{lehur2002} and be experimentally more challenging than far from charge degeneracy.

\section{Conclusion}\label{sec:conclusion}

In summary, we have demonstrated that the transport behaviour of a topological Kondo box hosting Majorana bound states depends sensitively on the proximity to a charge degeneracy point. Away from charge degeneracy, the box exhibits the well-studied topological Kondo effect, flowing to  maximum conductance for non-interacting leads, and displaying a quantum phase transition between strong-coupling and insulating regimes when the leads form Luttinger liquids. The situation is markedly different when the system is tuned to charge degeneracy. It also depends on the number of leads connected to the topological box and we can identify two situations:
\begin{enumerate}
\item For three or four Luttinger liquid leads, the intermediate Kondo fixed point is stable and exists for $\frac{M-1}{2 M} < K < \frac{M}{2 (M-1)}$, connecting the weak and strong coupling limits. As a result, the conductance between the leads takes the form
\begin{equation}
 G_{j,k}=\frac{2e^2 K \mu (K)}{h}\left(\frac{1}{M} -\delta_{j,k}\right),
 \end{equation} 
where the mobility $\mu (K)$, shown in Fig.~\ref{fig:PD2_Box}, varies between $0$ and $1$. We recover the multi-channel Kondo fixed point at $K=1/2$, with the intermediate mobility $\mu(1/2) = \sin^2[\pi/(M+2)]$.
\item For more than four leads, we recover the critical behaviour of the topological Kondo model at high mobility, which indicates that the pseudo-spin does not play a role if the Majorana fermions are already strongly coupled to the leads. A quantum phase transition thus occurs with maximum conductance on one side. The location of the transition is perturbatively the same as in the non-degenerate case, improving with the number of channels $M$. The difference at degeneracy is that the insulating phase is replaced by a weakly transmitting phase where the conductance decreases with $M$. This weak coupling regime is in fact analytically connected to a special point at $K=1/2$ where we recover exactly the multi-channel Kondo model.
\end{enumerate}
Our QBM analysis has identified the pseudo-spin, representing the two-state charge degeneracy in the box, as the physical ingredient explaining the difference in behaviour between the charge degenerate and non-degenerate cases. Detuning the system away from charge degeneracy or increasing the number of channels weakens the pseudo-spin component and thus extends topological Kondo physics in the phase diagram. Moreover, the intermediate fixed point $(I)$ is not robust against channel asymmetries, and therefore requires fine-tuning, in contrast with the strong coupling fixed point as discussed in Ref.~\onlinecite{Michaeli2016}.

We note that the point $K = K_C (M)$ at which the critical line $(I)$ turns over is not known analytically and remains a conjecture. A more precise study could be performed at large $M$ where the change of sign of $C_M (K)$ with $K$ should determine the location of $K_C$. Numerical results~\cite{Galpin-2014} would also nicely complement our work by discussing the phase diagram more quantitatively, in particular for $M\geq 5$  with Fig. \ref{fig:PD2_Box}.


\textit{Acknowledgements:} This work has benefited from useful discussions with
R. Egger. We acknowledge financial support from the PALM Labex, Paris-Saclay, Grant No. ANR-10-LABX-0039 and by the German Science Foundation (DFG) FOR2414. We also acknowledge discussions at CIFAR meetings in Canada, and conferences in Trieste
and in Mainz. 

\begin{appendix}
\section{Tunneling term}\label{App:Tunneling}
In this appendix, we detail a rigorous derivation of the tunneling Hamiltonian \eqref{eq:tunnel}.

First we consider that each lead is coupled to the extremity of a different nanowire, leading to the existence of at least $2M$ Majoranas. $N$ (even) is the total number of Majorana  zero modes  on the island.  Let $\gamma_{L/R,j}$ be the Majorana fermions at the extremity of each nanowire, and $d_j=\frac{1}{2} \left(\gamma_{L,j} + i\gamma_{R,j}\right)$ the corresponding delocalized fermion. The tunneling from the lead to the superconducting island can be written:
\begin{equation}
-t_j (d^\dagger_j + d_j e^{2i \chi} )\psi_j(0) + h.c.
\end{equation}
While the first term is the usual hopping term of a single fermion, the second one describes an alternative process where an electron of the lead and an electron on the nanowire combine to create a Cooper pair in the bulk of the superconducting island.

We wish to define a new operator $e^{i\tilde{\chi}}$, verifying 
\begin{equation}\label{eq:comm}
[\hat{N}, e^{i\tilde{\chi}} ]= e^{i\tilde{\chi}},
\end{equation}
where $\hat{N}=2\hat{N}_C + \hat{n}$ is the total number of particles in the box. Let us emphasize an important idea. While the proof of the existence of an actual operator $\tilde{\chi}$ verifying these properties is not trivial and in fact not even well-defined, the construction of $e^{i\tilde{\chi}}$ is much simpler and sufficient in practice, in analogy with the phase of a condensate. Similarly, $e^{2 i \chi}$ is well-defined operator, adding one Cooper pair to the island, but $e^{i \chi}$ and $\chi$ are not. Therefore $e^{i \chi}$ and $e^{i\tilde{\chi}}$ can not be blindly identified since the first one is ill-defined.

One can actually explicitly construct the operator $e^{i\tilde{\chi}}$ with the required properties
\begin{align}
e^{i \tilde{\chi}} &= (d^\dagger_1 + d_1 e^{2i \chi})e^{i\pi \sum\limits_{j=2}^{N/2} d^\dagger_j d_j}\times P_\text{leads}\\
&=(d^\dagger_1 -d_1 e^{2i \chi})\times P_\text{tot}
\end{align}
where $P_\text{leads}$ is the fermionic parity in all leads and $P_\text{tot}$ the total fermionic parity. Their only use is to ensure commutation with all fermionic operators in the leads. $e^{i \tilde{\chi}}$ is unitary (in the limit where $\hat{N}_c \gg 1$) and verifies the desired commutation relation~\eqref{eq:comm}. From its expression, one checks that $e^{i\tilde{\chi}} e^{-2i\chi}=e^{-i \tilde{\chi}}$ and $(e^{i\tilde{\chi}})^2=e^{2i \chi}$. In addition,
\begin{equation}\label{eq:comm2}
[e^{i \tilde{\chi}}, d_j]=[e^{i \tilde{\chi}}, d_j^\dagger]=0~~~\forall j>1 \\
\end{equation}
Hence, $e^{i\tilde{\chi}}$ satisfies all properties expected for the ill-defined operato $e^{i \chi}$ and it becomes legitimate to identify them in the model.

From Eq.~\eqref{eq:comm2}, the tunneling term can be rewritten as
\begin{equation}
-t_j \gamma_j e^{i\tilde{\chi}} \psi_j(0) + h.c.,
\end{equation}
where
\begin{align}
\gamma_j&=d^\dagger_j e^{-i \tilde{\chi}} + d_j e^{i \tilde{\chi}}\qquad \forall j>1\\
\gamma_1&=e^{i\pi \sum\limits_{j=2}^{N/2} d^\dagger_j d_j}\times P_\text{leads}.
\end{align}
The $\gamma_j$ operators all commute with $\hat{N}$ (and with the total number of fermions in general), and consequently totally decouples from the charge sector. They can essentially be understood as involutions mapping the $2^{N/2-1}$-dimensional subspace of the states $\prod_{j=1}^{N/2} d^{\dagger \nu_j}_j \Ket{0}, ~~\nu_j=0/1$ with even number of fermions on the one with an odd number of fermions, while modifying the number of Cooper pairs so that the total number of electrons in the box is conserved. Finally, they are indeed Majorana fermions as they are Hermitians and verify the Clifford algebra
\begin{equation}
\{\gamma_j, \gamma_k\}=2\delta_{j,k}
\end{equation}
They also anti-commute with all fermionic operators in the leads. Using the controlled identification $e^{i\tilde{\chi}}\rightarrow e^{i \chi}$, we indeed  recover the Hamiltonian \eqref{eq:tunnel}.

This proof can be generalized to the case where we attach leads at both extremities of wires, as long as $N>M$. To do so, we choose the following convention for the tunneling term:
\begin{multline}
-t_{2j}(d^\dagger_j+e^{2i\chi} d_j)\psi_{2j}(0)\\
-t_{2j-1}i(d^\dagger_j-e^{2i\chi} d_j)\psi_{2j-1}(0)  +h.c. ~~\forall j\geq 1,
\end{multline}
with $t_1=0$ (\textit{i.e.} we choose that an uncoupled Majorana is in the lead $1$). Then, the expression for the operator $e^{i\tilde{\chi}}$ is unchanged.

For $N=M$, one can no longer find enough independent hermitian matrices in the $2^{M/2-1}$-dimensional subspace and the $\gamma$ matrices no longer verify the Clifford algebra: following the previous convention, $\gamma_1$ and $\gamma_2$ commute. 

\section{Convention for the QBM}\label{App:Rmatrix}
The transformation used to decouple the total mode $\frac{1}{\sqrt{M}} \sum\limits_{j} \theta_j$ must satisfy two criteria. First, it must be an orthogonal transformation to respect the commutation relations of the bosonic fields. Second, it must generate the total mode within the new coordinates. For M leads, a convenient matrix $R$ that we shall use is given by
\begin{align}
R_{i,j}&= 0, ~\forall ~i>j+1 ~~~~ R_{i,j} = \frac{1}{\sqrt{i(i+1)}}, ~\forall ~i\le j<M \\
R_{i,i+1}&= \frac{-i}{\sqrt{i(i+1)}}, ~\forall i<M ~~~~ R_{M,i}=\frac{1}{\sqrt{M}} 
\end{align}
While all matrices $R$ lead to the same result, this choice simplify some evaluations. The vectors $\vec{w}_j$ can directly be read in $R$
\begin{equation}
\vec{w}_j=(R_{1,j}, ..., R_{M-1,j})
\end{equation}

\section{RG flow}

In this appendix, we recall the main steps of the derivation the RG equations for the M-CKM and the Topological Kondo model. To do so, we use a systematic expansion of the partition function, following Ref.~\onlinecite{Yi-2002}.

\subsection{RG equations for the Topological Kondo model}\label{App:RGTK}
\subsubsection{Weak coupling}
Let us start with the easier case of the Topological Kondo model.
We first consider the topological Kondo model corresponding to the absence of charge degeneracy. We use the QBM formalism for simplicity. We define the vectors $\vec{w}_j$ such that $\theta_j=\vec{w}_j. \vec{r} + \frac{1}{\sqrt{M}} \Theta$. They verify $\vec{w}_j. \vec{w}_k=\delta_{j,k}-\frac{1}{M}$. Explicit expressions can be found in appendix \ref{App:Rmatrix}. The action governing the model at weak coupling is
\begin{equation}
\begin{split}
S =  \sum_{\omega_m} \sum_{j}^{M-1} & \frac{|\omega_m| K}{2 \pi \beta }|r_{j} (\omega_m)|^2 \\
& -\sum\limits_{j \neq k}^{M,M} \lambda_{j,k} \int_0^\beta \frac{d \tau}{\tau_c}  \cos((\vec{w}_j-\vec{w}_k).\vec{r}),
\end{split}
\end{equation}
where $\tau_c$ is a short distance imaginary time cut-off. One can then proceed to a systematic expansion of the partition function using $\lambda_{j,k}$ as a small parameter. At order n, we obtain the contribution
\begin{multline}
   \int\limits_{\tau_1 < ..<\tau_l} \Braket{\prod\limits_{l=1}^n \frac{d \tau_l}{\tau_c}  \sum\limits_{j_l \neq k_l}^{M,M}
\lambda_{j_l,k_l} \cos((\vec{w}_{j_l}-\vec{w}_{k_l}).\vec{r}(\tau_l))}_0,
\end{multline}
where $\Braket{...}_0$ is the average value with the unperturbed action. We proceed then to real-space renormalization in imaginary time, i.e. we increase the cut-off $\tau_c$ to $\tau_c'=\tau_c e^{\ell}$, and fuse operators closer than $\tau_c'$. Using the invariance of the action upon renormalization of the cut-off, we derive the different RG equations. We recall the correlation functions for the free bosons (we include an infrared cut-off in the integrals):
\begin{equation}
\Braket{\prod\limits_j e^{i \vec{w}_{k_j}.\vec{r}(\tau_j)}}_0= \delta_{\sum\limits_j \vec{w}_{k_j}=0}\prod\limits_{j<l} \left(\frac{\tau_c^2}{(\tau_j-\tau_l)^2} \right)^{-\frac{\vec{w}_{k_j}.\vec{w}_{k_l} }{2K}} \label{eq:corrfunc}
\end{equation}
A first contribution  is generic: when $\tau_2-\tau_1\gg \tau_c'$, 
\begin{align*}
&\lambda_{j,k} \lambda_{k,j} \int\limits_{\tau_1 < \tau_2} \frac{d \tau_1 d\tau_2}{\tau_c^2} 
 \\
 &\qquad \qquad\Braket{\cos((\vec{w}_j-\vec{w}_k).\vec{r}(\tau_1))\cos((\vec{w}_k-\vec{w}_j).\vec{r}(\tau_2))}_0 \\
 &=\lambda_{j,k} \lambda_{k,j} \int\limits_{\tau_1 < \tau_2} \frac{d \tau_1 d\tau_2}{\tau_c^2} \left( \frac{\tau_c^2}{(\tau_1-\tau_2)^2} \right)^{\frac{1}{K}} 
 \end{align*}
This last expression must be cut-off independent, imposing $\lambda_{j,k}(\ell) \lambda_{k,j}(\ell)=e^{(2-\frac{2}{K}) \ell}\lambda_{j,k} \lambda_{k,j}$, or by symmetry,  $\lambda_{j,k}(\ell)=e^{(1-\frac{1}{K}) \ell}\lambda_{j,k} $\\
When $\tau_2-\tau_1 < \tau_c'$, the operators are no longer taken separately and fuse together. In particular, if $j_1=k_2, k_1\neq j_2$, we generate additional terms $\cos((\vec{w}_{j_2}-\vec{w}_{k_1}).\vec{r}(\tau_1))$ (alternatively for $j_2=k_1, k_2\neq j_1$). To rigorously compute the exact coefficient for this contribution, we use the third order terms $\lambda_{j_1, k_1}\lambda_{k_1, k_2}\lambda_{k_2, j_1}$, with $\tau_3-\tau_1 \gg \tau_c'$. We need to evaluate, for $j\neq k \neq l \neq j$:
\begin{align}
&\int\limits_{\tau_c<\tau<\tau_c e^{\ell}} \Braket{e^{i(w_{j}-w_{k}).\vec{r}(\tau_1)} e^{i(w_{k}-w_{l}).\vec{r}(\tau_1+\tau)} e^{i(w_{l}-w_{j}).\vec{r}(\tau_3)}}_0\\
&=\left(\frac{\tau_c^2}{(\tau_3-\tau_1)^2} \right)^{\frac{1}{2K}} \int\limits_{\tau_c<\tau<\tau_c e^{\ell}} \left(\frac{\tau_c^2}{\tau^2} \frac{\tau_c^2}{(\tau_3-\tau_1-\tau)^2} \right)^{\frac{1}{2K}} \\
&\left(\frac{\tau_c^2}{(\tau_3-\tau_1)^2} \right)^{\frac{1}{K}} \ell \qquad \text{if $\ell\ll 1$} \label{eq:simplecase}
\end{align}
We finally obtain for $\ell \ll 1$, 
$$\lambda_{j,k} (\ell) = \lambda_{j,k}+ (1-\frac{1}{K}) \ell \lambda_{j,k} + 2 \ell \sum\limits_{m \neq j, k} \lambda_{j,m} \lambda_{m,k},$$ leading to the RG equation~\eqref{eq:RGTK}. 

\subsubsection{Strong coupling}

At strong coupling, the action describing instanton excitations is given by Eq.~\eqref{eq:strong} reproduced here,
\begin{equation}\label{eq:strong2}
S =  \sum_{\omega_m} \sum_{j}  \frac{|\omega_m| }{2 \pi K \beta} |\phi_{j} (\omega_m)|^2 
 - v \int_0^\beta d \tau \sum_k \hat{O}_k (\tau), 
\end{equation}
with the operators $\hat{O}_k=e^{ 2i (\phi_k-\frac{1}{\sqrt{M}} \Phi ) }$ connecting the lattice of minima. Proceeding with an expansion in powers of $v$, the calculation is similar as at weak coupling except for the replacement $K\rightarrow \frac{1}{K}$.

For $M=3$, we note that $\hat{O}_1 \hat{O}_2 = \hat{O}^\dagger_3$. The computation of the third order coefficient is identical to Eq.~\eqref{eq:simplecase} and the RG equation is given by
\begin{equation}\label{eq:rg1}
\frac{dv}{d\ell} =  \left(1-\frac{4 K}{3} \right)v+2v^2
\end{equation}
For $M>3$, the first non-zero terms are at third order in $v$. The two types contributions one has to take into account are: $\hat{O}_1 \hat{O}_2 \hat{O}_2^\dagger = \hat{O}_1$ and, for $M=4$ only, $\hat{O}_1 \hat{O}_2 \hat{O}_3= \hat{O}^\dagger_4$. The computation of the corresponding terms lead to the RG equation~\cite{Yi-2002}
\begin{equation}\label{eq:rg2}
\frac{dv}{d\ell} =  \left(1-\frac{2 (M-1)K}{M} \right)v + (B_M + A_4 \delta_{M,4}) v^3
\end{equation}
where $B_M$ is a positive coefficient, see also Eq.~\eqref{eq:a3}. Eqs.~\eqref{eq:rg1} and~\eqref{eq:rg2} predict that for all $M \ge 3$, the strong coupling fixed point is unstable for $K \le \frac{M}{2(M-1)}$, while for $K > \frac{M}{2(M-1)}$, an unstable fixed point $(I)$ emerges at finite $v$.

\subsection{RG equations for the M-CKM}\label{App:RG-MCKM}
\subsubsection{Weak coupling}

The Hamiltonian obtained after the unitary transformation~\eqref{eq:unitary} is given  Eq.~\eqref{eq:boso}, and can also be written as 
\begin{multline}
\hat{U}^\dagger H \hat{U}=H_\text{leads}\{\tilde{\phi}, \tilde{\theta}, \tilde{K}\} \\
- \sum\limits_{j=1}^{M} \left( \frac{J_{\perp, j}}{2} \tau_+ e^{i \sqrt{2} \vec{w}_j .\tilde{r}_j+\sqrt{2}\left(\frac{1}{\sqrt{M}}-\frac{\tilde{K} J_z \sqrt{M}}{2} \right)\tilde{\Theta}} +h.c \right).
\end{multline}
We introduce the notation $M_z=\frac{1}{\sqrt{M}}-\frac{\tilde{K} J_z \sqrt{M}}{2}$ and compute the RG equations for both $J_\perp$ and $M_z$.

We proceed with a similar expansion of the partition function using $J_\perp$ as a small parameter. The contribution of order $n$ is
\begin{equation}
\left( \frac{J_\perp}{2} \right)^n \int\limits_{\tau_1<...<\tau_n} \frac{d\tau_1... d\tau_n}{\tau_c^n} \sum\limits_{\rm n-loops} \exp\left(\sum\limits_{j<k}^{n,n} V_{j,k}\right) + h.c.,
\end{equation}
where $\sum\limits_{\rm n-loops}$ signify that we sum over all n-uplets $(a_j)$ such that $\sum\limits (-1)^j \vec{w}_{a_j} = 0$ and (using Eq. \eqref{eq:corrfunc})
\begin{equation}
V_{j,k}=2\frac{(-1)^{j+k}}{\tilde{K}}\left(M_z^2 + \vec{w}_{a_j}.\vec{w}_{a_k} \right) \log(\frac{\tau_k-\tau_j}{\tau_c})
\end{equation}
The alternating signs take into account the spin operators. Similarly, we increase the cut-off $\tau_c$ to $\tau_c'=\tau_c e^{\ell}$, and fuse the operators when needed. To lowest order, only two consecutive operators can fuse. The most relevant contribution appear when these two contributions have the same $\vec{w}_{a_j}$. Let us assume that this happens for the $l^\text{th}$ and $l^\text{th}+1$ operators and define $\tau=\tau_{l+1}-\tau_l<\tau_c'$ and
\begin{align*}
& V_d(\tau_l, \tau, a_l)=\sum\limits_{j<l} V_{j,l}+V_{j,l+1} + \sum\limits_{j>l+1} V_{l,j}+V_{l+1,j}\\
&\approx 2\sum\limits_{j\neq l, l+1} \frac{(-1)^{j+l}}{\tilde{K}}\left(M_z^2 + \vec{w}_{a_j}.\vec{w}_{a_l} \right) \tau \partial_{\tau_l} \log(\frac{|\tau_l-\tau_j|}{\tau_c}) 
\end{align*}
Integrating over the two variables $\tau_l$ and $\tau$ reduces the n-loop to a n-2 loop and consequently, at order n, we have an additional contribution coming from the order n+2,
\begin{align*}
&\left(\frac{J_\perp}{2}\right)^2 \sum\limits_{j=0}^{n} \sum\limits_a \int\limits_{\tau_j}^{\tau_{j+1}}\frac{d\tau'}{\tau_c}\int\limits_{\tau_c}^{\tau_c e^\ell} \frac{d\tau}{\tau_c} e^{V_d(\tau', \tau, a)} \\
&\approx \left(\frac{J_\perp}{2} \right)^2 \sum\limits_{j=0}^{n} \sum\limits_a \int\limits_{\tau_j}^{\tau_{j+1}}\frac{d\tau'}{\tau_c}\int\limits_{\tau_c}^{\tau_c e^\ell} \frac{d\tau}{\tau_c} 1+ V_d(\tau', \tau, a)\\
&\approx \left(\frac{J_\perp}{2}\right)^2 \left( \frac{M\beta(e^\ell-1)}{\tau_c}\right.
\\
&\qquad \qquad-8M \left.\frac{M_z^2(e^\ell-1)}{\tilde{K}} \sum\limits_{j<k}(-1)^{j+k}	 \log(\frac{\tau_k-\tau_j}{\tau_c}) \right),
\end{align*}
where $\tau_0=0$ and $\tau_{n+1}=\beta$. While the first term can be ignored, as it corresponds to a rescaling of the ground state energy, the second term indeed renormalizes the partition function. Reexponentiation leads to a correction of $V_{p,q}$ given by
\begin{equation}
V_{p,q}\rightarrow V_{p,q} - \frac{M J_\perp^2 (e^\ell-1)}{K} M^2_z (-1)^{p+q} \log(\frac{\tau_q-\tau_p}{\tau_c})
\end{equation}
 For small $\ell$, we have
\begin{equation}
M_z(\ell)^2=M_z^2-M J_\perp^2 M_z^2 \ell,
\end{equation}
or
\begin{equation}
\frac{d J_z}{dl} = J_\perp^2 \left(\frac{1}{\tilde{K}}-\frac{M J_z}{2} \right),
\end{equation}
which is the first RG equation.
To obtain the RG equations for $J_\perp$, we simply rescale $\tau_c$ in both the integrals and $V_{j,k}$ and extract the $n$ dependency. We obtain
\begin{align*}
J_\perp^n&=J_\perp^n(\ell)e^{-n\ell} e^{-2\ell \sum\limits_{j<k}\frac{(-1)^{j+k}}{\tilde{K}}( M_z^2+\vec{w}_{a_j}.\vec{w}_{a_k})}\\
&=J_\perp^n(\ell) e^{-n\ell} e^{\frac{n}{\tilde{K}}\ell ( M_z^2+\frac{M-1}{M})}\\
\frac{dJ_\perp}{d\ell}&=\left(1-\frac{M-1}{M \tilde{K}}-\frac{M_z^2}{\tilde{K}}\right)J_\perp\\
&=\left(1-\frac{1}{ \tilde{K}}+J_z[1-\frac{\tilde{K} J_z M}{4}]\right)J_\perp
\end{align*}

\subsubsection{Strong coupling}

In this case, the vicinity of the strong coupling fixed point $(S)$ is governed by the action
\begin{equation}\label{eq:strong3}
S =  \sum_{\omega_m} \sum_{j}  \frac{|\omega_m| }{2 \pi K \beta} |\phi_{j} (\omega_m)|^2 
 - v \int_0^\beta d \tau \sum_k \hat{O}_k^{(h)} (\tau), 
\end{equation}
where the $\hat{O}_k^{(h)}$ operators, with dimension $\frac{2K (M-1)}{M}$, are given in Eq.~\eqref{eq:trans}. The difference with the operators $\hat{O}_k$ encountered in the topological Kondo model, see Eq.~\eqref{eq:strong2}, is in the pseudo-spin rotation $e^{-i \pi \tau_z/M}$.

For $M=3$, $\hat{O}_1 \hat{O}_2 = -\hat{O}^\dagger_3$, where the minus sign stems from the Berry phase of the spin wavefunction, $e^{-i\pi \tau_z}=-1$. The calculation of the RG equation is this almost identical to \eqref{eq:simplecase}, 
\begin{equation}
\frac{dv_2}{d\ell} = (1-4 K/3)v-2v^2
\end{equation}
except for the sign change in the last term. In general for $M > 3$, the first non-zero contribution to the RG equation (apart from a linear term) is third order in $v$. For $M=4$, there is a competition between two contributions, the contraction $\hat{O}_1 \hat{O}_2 \hat{O}_2^\dagger = \hat{O}_1$ in which the spin plays no role, leading to the coefficient
\begin{equation}
\begin{split}
B_4 = 6 \int_0^1 d x   & \left [ \frac{x^{2/3} + x^{-2/3}-2}{(1-x)^2} 
\right. \\
& \left. + \left( \frac{x}{1-x}  \right)^{2/3} -1 \right ] 
\end{split}
\end{equation}
and the contribution $\hat{O}_1 \hat{O}_2 \hat{O}_3= -\hat{O}^\dagger_4$, carrying the spin Berry phase, associated to the coefficient
\begin{equation}\label{eq:a3}
A_4 = 6  \int\limits_0^1 d x \frac{1}{x^{2/3} (1-x)^{2/3} }
\end{equation}
The RG equation takes the form 
\begin{equation}
\frac{dv_3}{d\ell} =  \left(1-\frac{3 K}{2}\right)v-(A_4-B_4)v^3
\end{equation}
with $A_4 - B_4>0$ such that the spin wavefunction eventually governs the transition. We obtain a phase diagram similar to $M=4$ as explained in the main text. 

For $M\ge 5$, only $\hat{O}_1 \hat{O}_2 \hat{O}_2^\dagger = \hat{O}_1$ contributes to third order in $v$. The RG equation takes the form~\eqref{eq:RG5} in the main text, exactly the same as in the charge non-degenerate case Eq.~\eqref{eq:rg2}, and the phase diagram differs from $M \le 4$.

\section{Kubo approach to conductance}\label{App:Kubo}
To compute the conductance in the Majorana island, we will use Kubo formula. We present in this Appendix a short derivation of the Kubo formula for our model, before an example of application far from charge degeneracy (see for example Ref \onlinecite{Galpin-2014} for an alternative derivation). We express the conductance as a correlation function of the initial bosonic field first, and then as a correlation function of the $(\vec{k}, \Phi)$ fields.

\subsection{Kubo formula}
We start from linear response theory.   Given a small perturbation $H'(t)$ switched on adiabatically, the change for the average value of the observable $A$ is:
\begin{equation}
\Braket{\Delta A(t)}=\frac{-i}{\hbar }  \int\limits_{-\infty}^t e^{-\eta(t- t')}\Braket{[A(t), H'(t')]}dt'
\end{equation}
with $\eta \rightarrow 0^+ $. Let $C_{A,B}^R(t-t')=-\frac{i}{\hbar}\theta(t-t') \Braket{[A(t), B(t')]}$. To compute $C_{A,B}^R(\omega)$, we compute another correlation function and do an analytic continuation. Let $C_{A,B}(\tau)=-\Braket{T_\tau A(\tau) B(0)}$. Starting with:
\begin{align}
A(\tau) &= \frac{1}{\beta} \sum\limits_{n=- \infty}^{n=+\infty} e^{-i\omega_n \tau}A(i\omega_n )\\
B(0) &= \frac{1}{\beta} \sum\limits_{n=- \infty}^{n=+\infty} B(i\omega_n )\\
\end{align}
we obtain:
\begin{align}
C_{A,B}(i\omega_n) &=-\int\limits_0^\beta d\tau e^{i\omega_n \tau} \frac{1}{\beta^2} \sum\limits_{m,k} e^{-i\omega_n \tau} \Braket{A(i\omega_m) B(i\omega_k)} \\
C_{A,B}(i\omega_n) &= -\frac{1}{\beta} \sum\limits_k \Braket{A(i\omega_n) B(i \omega_k)}
\end{align}
In particular,
\begin{equation}
C_{\dot{A},B}(i\omega_n) = -\frac{1}{\beta} \sum\limits_k i\omega_n \Braket{A(i\omega_n) B(i \omega_k)}
\end{equation}
The current operator for the wire $j$ is given by $e \partial_t N_j$, where $N_j$ is the total charge in the wire. The linear conductance $G_{j,k}$, corresponding to the current in the $j^\text{th}$ wire due to a potential in the $k^\text{th}$ wire is $\frac{\partial \Braket{\Delta e \partial_t N_j}}{\partial V_k}$, where $V_k$ is the potential in the wire $k$.
Given $A=e N_j$ and $B=e V_k N_k$, we finally obtain the conductance $G(\omega)$:
\begin{equation}
G_{j,k}(\omega) =-\frac{e^2}{h} \int d\nu \omega \Braket{N_j(\omega) N_k(\nu)}
\end{equation}
Finally, as we are interested in the DC conductance, we obtain:
\begin{equation}
G=-\frac{e^2}{h} \lim\limits_{\omega \rightarrow 0}  \int d\nu \omega \Braket{N_j(\omega) N_k(\nu)}
\end{equation}
As our action is will be diagonal in the Matsubara frequencies, it simplifies to:
\begin{equation}
G_{j,k}=-\frac{e^2}{h} \lim\limits_{\omega \rightarrow 0} \omega \Braket{N_j(\omega) N_k(\omega)}
\end{equation}

\subsection{Application for the strong coupling limit far from charge degeneracy}

For semi-infinite LL wires, $N=\frac{\phi(x=0)}{\pi}$. We want to express $N_j$ as a function of $(\vec{k}, \Phi)$, dual to $(\vec{r}, \Theta)$. 
\begin{align}
G_{j,k}&=-\frac{e^2}{\pi^2 h}\lim\limits_{\omega \rightarrow 0} \omega \Braket{\phi_j (i \omega) \phi_k(i \omega)} \\
G_{j,k}&=-\frac{e^2}{\pi^2 h} \sum\limits_{l,m}^{M-1} \vec{w}_k(l)\vec{w}_j(m) \lim\limits_{\omega \rightarrow 0} \omega \Braket{k_l (i \omega) k_m(i\omega)}, 
\end{align}
where $\vec{w}_k(l)$ is the $l^\text{th}$ component of $\vec{w}_k$. The global mode does not intervene as it is pinned due to charge conservation, and consequently any contribution vanish. $\vec{k}$ being the dual of $\vec{r}$, when the latter are pinned in the strong coupling limit, $\vec{k}$ is free and one obtains:
\begin{equation}
\lim\limits_{\omega \rightarrow 0} \omega  \Braket{k_l (i \omega) k_m(i \omega)} =2 \pi^2 K \delta_{l,m},
\end{equation}
leading to the celebrated conductance:
\begin{align}
G_{k,j}&=-\frac{2Ke^2}{h} \vec{w}_{k}.\vec{w}_j \\
&=\frac{2Ke^2}{h} (\frac{1}{M}-\delta_{k,j})
\end{align}

\end{appendix}

\bibliography{Paper_Kondo_6}

\begin{thebibliography}{60}%
\makeatletter
\providecommand \@ifxundefined [1]{%
 \@ifx{#1\undefined}
}%
\providecommand \@ifnum [1]{%
 \ifnum #1\expandafter \@firstoftwo
 \else \expandafter \@secondoftwo
 \fi
}%
\providecommand \@ifx [1]{%
 \ifx #1\expandafter \@firstoftwo
 \else \expandafter \@secondoftwo
 \fi
}%
\providecommand \natexlab [1]{#1}%
\providecommand \enquote  [1]{``#1''}%
\providecommand \bibnamefont  [1]{#1}%
\providecommand \bibfnamefont [1]{#1}%
\providecommand \citenamefont [1]{#1}%
\providecommand \href@noop [0]{\@secondoftwo}%
\providecommand \href [0]{\begingroup \@sanitize@url \@href}%
\providecommand \@href[1]{\@@startlink{#1}\@@href}%
\providecommand \@@href[1]{\endgroup#1\@@endlink}%
\providecommand \@sanitize@url [0]{\catcode `\\12\catcode `\$12\catcode
  `\&12\catcode `\#12\catcode `\^12\catcode `\_12\catcode `\%12\relax}%
\providecommand \@@startlink[1]{}%
\providecommand \@@endlink[0]{}%
\providecommand \url  [0]{\begingroup\@sanitize@url \@url }%
\providecommand \@url [1]{\endgroup\@href {#1}{\urlprefix }}%
\providecommand \urlprefix  [0]{URL }%
\providecommand \Eprint [0]{\href }%
\providecommand \doibase [0]{http://dx.doi.org/}%
\providecommand \selectlanguage [0]{\@gobble}%
\providecommand \bibinfo  [0]{\@secondoftwo}%
\providecommand \bibfield  [0]{\@secondoftwo}%
\providecommand \translation [1]{[#1]}%
\providecommand \BibitemOpen [0]{}%
\providecommand \bibitemStop [0]{}%
\providecommand \bibitemNoStop [0]{.\EOS\space}%
\providecommand \EOS [0]{\spacefactor3000\relax}%
\providecommand \BibitemShut  [1]{\csname bibitem#1\endcsname}%
\let\auto@bib@innerbib\@empty
\bibitem [{\citenamefont {Alicea}(2012)}]{alicea2012}%
  \BibitemOpen
  \bibfield  {author} {\bibinfo {author} {\bibfnamefont {J.}~\bibnamefont
  {Alicea}},\ }\bibfield  {title} {\enquote {\bibinfo {title} {New directions
  in the pursuit of majorana fermions in solid state systems},}\ }\href
  {http://iopscience.iop.org/article/10.1088/0034-4885/75/7/076501/meta}
  {\bibfield  {journal} {\bibinfo  {journal} {Rep. Prog. Phys. 75, 076501}\ }
  (\bibinfo {year} {2012})}\BibitemShut {NoStop}%
\bibitem [{\citenamefont {Beenakker}(2013)}]{Beenakker2013}%
  \BibitemOpen
  \bibfield  {author} {\bibinfo {author} {\bibfnamefont {C.~W.~J.}\
  \bibnamefont {Beenakker}},\ }\bibfield  {title} {\enquote {\bibinfo {title}
  {Search for majorana fermions in superconductors},}\ }\href {\doibase
  10.1146/annurev-conmatphys-030212-184337} {\bibfield  {journal} {\bibinfo
  {journal} {Annual Review of Condensed Matter Physics}\ }\textbf {\bibinfo
  {volume} {4}},\ \bibinfo {pages} {113--136} (\bibinfo {year}
  {2013})}\BibitemShut {NoStop}%
\bibitem [{\citenamefont {Read}\ and\ \citenamefont {Green}(2000)}]{Read2000}%
  \BibitemOpen
  \bibfield  {author} {\bibinfo {author} {\bibfnamefont {N.}~\bibnamefont
  {Read}}\ and\ \bibinfo {author} {\bibfnamefont {D.}~\bibnamefont {Green}},\
  }\bibfield  {title} {\enquote {\bibinfo {title} {Paired states of fermions in
  two dimensions with breaking of parity and time-reversal symmetries and the
  fractional quantum hall effect},}\ }\href {\doibase
  10.1103/PhysRevB.61.10267} {\bibfield  {journal} {\bibinfo  {journal} {Phys.
  Rev. B}\ }\textbf {\bibinfo {volume} {61}},\ \bibinfo {pages} {10267--10297}
  (\bibinfo {year} {2000})}\BibitemShut {NoStop}%
\bibitem [{\citenamefont {Kitaev}(2001)}]{Kitaev2001}%
  \BibitemOpen
  \bibfield  {author} {\bibinfo {author} {\bibfnamefont {A.}~\bibnamefont
  {Kitaev}},\ }\bibfield  {title} {\enquote {\bibinfo {title} {Unpaired
  majorana fermions in quantum wires},}\ }\href {\doibase
  10.1070/1063-7869/44/10S/S29} {\bibfield  {journal} {\bibinfo  {journal}
  {Physics Uspekhi}\ }\textbf {\bibinfo {volume} {44}},\ \bibinfo {pages} {131}
  (\bibinfo {year} {2001})}\BibitemShut {NoStop}%
\bibitem [{\citenamefont {{Das}}\ \emph {et~al.}(2012)\citenamefont {{Das}},
  \citenamefont {{Ronen}}, \citenamefont {{Most}}, \citenamefont {{Oreg}},
  \citenamefont {{Heiblum}},\ and\ \citenamefont {{Shtrikman}}}]{A.Das2012}%
  \BibitemOpen
  \bibfield  {author} {\bibinfo {author} {\bibfnamefont {A.}~\bibnamefont
  {{Das}}}, \bibinfo {author} {\bibfnamefont {Y.}~\bibnamefont {{Ronen}}},
  \bibinfo {author} {\bibfnamefont {Y.}~\bibnamefont {{Most}}}, \bibinfo
  {author} {\bibfnamefont {Y.}~\bibnamefont {{Oreg}}}, \bibinfo {author}
  {\bibfnamefont {M.}~\bibnamefont {{Heiblum}}}, \ and\ \bibinfo {author}
  {\bibfnamefont {H.}~\bibnamefont {{Shtrikman}}},\ }\bibfield  {title}
  {\enquote {\bibinfo {title} {Zero-bias peaks and splitting in an al–inas
  nanowire topological superconductor as a signature of majorana fermions},}\
  }\href {\doibase 10.1038/nphys2479} {\bibfield  {journal} {\bibinfo
  {journal} {Nat. Phys}\ }\textbf {\bibinfo {volume} {8}},\ \bibinfo {pages}
  {887--895} (\bibinfo {year} {2012})}\BibitemShut {NoStop}%
\bibitem [{\citenamefont {Oreg}\ \emph {et~al.}(2010)\citenamefont {Oreg},
  \citenamefont {Refael},\ and\ \citenamefont {von Oppen}}]{wire1}%
  \BibitemOpen
  \bibfield  {author} {\bibinfo {author} {\bibfnamefont {Y.}~\bibnamefont
  {Oreg}}, \bibinfo {author} {\bibfnamefont {G.}~\bibnamefont {Refael}}, \ and\
  \bibinfo {author} {\bibfnamefont {F.}~\bibnamefont {von Oppen}},\ }\bibfield
  {title} {\enquote {\bibinfo {title} {Helical liquids and majorana bound
  states in quantum wires},}\ }\href
  {http://link.aps.org/doi/10.1103/PhysRevLett.105.177002} {\bibfield
  {journal} {\bibinfo  {journal} {Phys. Rev. Lett. 105, 177002}\ } (\bibinfo
  {year} {2010})}\BibitemShut {NoStop}%
\bibitem [{\citenamefont {Lutchyn}\ \emph {et~al.}(2010)\citenamefont
  {Lutchyn}, \citenamefont {Sau},\ and\ \citenamefont {Das~Sarma}}]{wire2}%
  \BibitemOpen
  \bibfield  {author} {\bibinfo {author} {\bibfnamefont {R.M.}\ \bibnamefont
  {Lutchyn}}, \bibinfo {author} {\bibfnamefont {J.D.}\ \bibnamefont {Sau}}, \
  and\ \bibinfo {author} {\bibfnamefont {S.}~\bibnamefont {Das~Sarma}},\
  }\bibfield  {title} {\enquote {\bibinfo {title} {Majorana fermions and a
  topological phase transition in semiconductor-superconductor
  heterostructures},}\ }\href
  {http://link.aps.org/doi/10.1103/PhysRevLett.105.077001} {\bibfield
  {journal} {\bibinfo  {journal} {Phys. Rev. Lett. 105, 077001}\ } (\bibinfo
  {year} {2010})}\BibitemShut {NoStop}%
\bibitem [{\citenamefont {{Albrecht}}\ \emph {et~al.}(2016)\citenamefont
  {{Albrecht}}, \citenamefont {{Higginbotham}}, \citenamefont {{Madsen}},
  \citenamefont {{Kuemmeth}}, \citenamefont {{Jespersen}}, \citenamefont
  {{Nyg{\aa}rd}}, \citenamefont {{Krogstrup}},\ and\ \citenamefont
  {{Marcus}}}]{Albrecht2016}%
  \BibitemOpen
  \bibfield  {author} {\bibinfo {author} {\bibfnamefont {S.~M.}\ \bibnamefont
  {{Albrecht}}}, \bibinfo {author} {\bibfnamefont {A.~P.}\ \bibnamefont
  {{Higginbotham}}}, \bibinfo {author} {\bibfnamefont {M.}~\bibnamefont
  {{Madsen}}}, \bibinfo {author} {\bibfnamefont {F.}~\bibnamefont
  {{Kuemmeth}}}, \bibinfo {author} {\bibfnamefont {T.~S.}\ \bibnamefont
  {{Jespersen}}}, \bibinfo {author} {\bibfnamefont {J.}~\bibnamefont
  {{Nyg{\aa}rd}}}, \bibinfo {author} {\bibfnamefont {P.}~\bibnamefont
  {{Krogstrup}}}, \ and\ \bibinfo {author} {\bibfnamefont {C.~M.}\ \bibnamefont
  {{Marcus}}},\ }\bibfield  {title} {\enquote {\bibinfo {title} {{Exponential
  protection of zero modes in Majorana islands}},}\ }\href
  {http://www.nature.com/nature/journal/v531/n7593/full/nature17162.html}
  {\bibfield  {journal} {\bibinfo  {journal} {\nat}\ }\textbf {\bibinfo
  {volume} {531}},\ \bibinfo {pages} {206--209} (\bibinfo {year}
  {2016})}\BibitemShut {NoStop}%
\bibitem [{\citenamefont {Halperin}\ \emph {et~al.}(2012)\citenamefont
  {Halperin}, \citenamefont {Oreg}, \citenamefont {Stern}, \citenamefont
  {Refael}, \citenamefont {Alicea},\ and\ \citenamefont {von
  Oppen}}]{Halperin2011}%
  \BibitemOpen
  \bibfield  {author} {\bibinfo {author} {\bibfnamefont {B.~I.}\ \bibnamefont
  {Halperin}}, \bibinfo {author} {\bibfnamefont {Y.}~\bibnamefont {Oreg}},
  \bibinfo {author} {\bibfnamefont {A.}~\bibnamefont {Stern}}, \bibinfo
  {author} {\bibfnamefont {G.}~\bibnamefont {Refael}}, \bibinfo {author}
  {\bibfnamefont {J.}~\bibnamefont {Alicea}}, \ and\ \bibinfo {author}
  {\bibfnamefont {F.}~\bibnamefont {von Oppen}},\ }\bibfield  {title} {\enquote
  {\bibinfo {title} {Adiabatic manipulations of majorana fermions in a
  three-dimensional network of quantum wires},}\ }\href
  {http://link.aps.org/doi/10.1103/PhysRevB.85.144501} {\bibfield  {journal}
  {\bibinfo  {journal} {Phys. Rev. B}\ }\textbf {\bibinfo {volume} {85}},\
  \bibinfo {pages} {144501} (\bibinfo {year} {2012})}\BibitemShut {NoStop}%
\bibitem [{\citenamefont {Matveev}(1990)}]{matveev1991b}%
  \BibitemOpen
  \bibfield  {author} {\bibinfo {author} {\bibfnamefont {K.~A.}\ \bibnamefont
  {Matveev}},\ }\bibfield  {title} {\enquote {\bibinfo {title} {Quantum
  fluctuations of the charge of a metal particle under the coulomb blockade
  conditions},}\ }\href {http://www.msd.anl.gov/personnel/matveev/docs/7.pdf}
  {\bibfield  {journal} {\bibinfo  {journal} {Zh. Eksp. Teor. Fiz.}\ }\textbf
  {\bibinfo {volume} {99}},\ \bibinfo {pages} {1598} (\bibinfo {year}
  {1990})},\ \bibinfo {note} {[{\it Sov. Phys. JETP} {\bf 72,} 892
  (1991)]}\BibitemShut {NoStop}%
\bibitem [{\citenamefont {Matveev}(1995)}]{matveev1995}%
  \BibitemOpen
  \bibfield  {author} {\bibinfo {author} {\bibfnamefont {K.~A.}\ \bibnamefont
  {Matveev}},\ }\bibfield  {title} {\enquote {\bibinfo {title} {Coulomb
  blockade at almost perfect transmission},}\ }\href {\doibase
  10.1103/PhysRevB.51.1743} {\bibfield  {journal} {\bibinfo  {journal} {Phys.
  Rev. B}\ }\textbf {\bibinfo {volume} {51}},\ \bibinfo {pages} {1743--1751}
  (\bibinfo {year} {1995})}\BibitemShut {NoStop}%
\bibitem [{\citenamefont {Emery}\ and\ \citenamefont
  {Kivelson}(1992)}]{Emery1992}%
  \BibitemOpen
  \bibfield  {author} {\bibinfo {author} {\bibfnamefont {V.~J.}\ \bibnamefont
  {Emery}}\ and\ \bibinfo {author} {\bibfnamefont {S.}~\bibnamefont
  {Kivelson}},\ }\bibfield  {title} {\enquote {\bibinfo {title} {Mapping of the
  two-channel kondo problem to a resonant-level model},}\ }\href {\doibase
  10.1103/PhysRevB.46.10812} {\bibfield  {journal} {\bibinfo  {journal} {Phys.
  Rev. B}\ }\textbf {\bibinfo {volume} {46}},\ \bibinfo {pages} {10812--10817}
  (\bibinfo {year} {1992})}\BibitemShut {NoStop}%
\bibitem [{\citenamefont {Clarke}\ \emph {et~al.}(1993)\citenamefont {Clarke},
  \citenamefont {Giamarchi},\ and\ \citenamefont {Shraiman}}]{Clarke1993}%
  \BibitemOpen
  \bibfield  {author} {\bibinfo {author} {\bibfnamefont {D.~G.}\ \bibnamefont
  {Clarke}}, \bibinfo {author} {\bibfnamefont {T.}~\bibnamefont {Giamarchi}}, \
  and\ \bibinfo {author} {\bibfnamefont {B.~I.}\ \bibnamefont {Shraiman}},\
  }\bibfield  {title} {\enquote {\bibinfo {title} {Curie and non-curie behavior
  of impurity spins in quantum antiferromagnets},}\ }\href
  {http://link.aps.org/doi/10.1103/PhysRevB.48.7070} {\bibfield  {journal}
  {\bibinfo  {journal} {Phys. Rev. B}\ }\textbf {\bibinfo {volume} {48}},\
  \bibinfo {pages} {7070--7076} (\bibinfo {year} {1993})}\BibitemShut {NoStop}%
\bibitem [{\citenamefont {Sengupta}\ and\ \citenamefont
  {Georges}(1994)}]{Sengupta1994}%
  \BibitemOpen
  \bibfield  {author} {\bibinfo {author} {\bibfnamefont {A.~M.}\ \bibnamefont
  {Sengupta}}\ and\ \bibinfo {author} {\bibfnamefont {A.}~\bibnamefont
  {Georges}},\ }\bibfield  {title} {\enquote {\bibinfo {title} {Emery-kivelson
  solution of the two-channel kondo problem},}\ }\href
  {http://link.aps.org/doi/10.1103/PhysRevB.49.10020} {\bibfield  {journal}
  {\bibinfo  {journal} {Phys. Rev. B}\ }\textbf {\bibinfo {volume} {49}},\
  \bibinfo {pages} {10020--10022} (\bibinfo {year} {1994})}\BibitemShut
  {NoStop}%
\bibitem [{\citenamefont {Coleman}\ \emph {et~al.}(1995)\citenamefont
  {Coleman}, \citenamefont {Ioffe},\ and\ \citenamefont
  {Tsvelik}}]{Coleman1995}%
  \BibitemOpen
  \bibfield  {author} {\bibinfo {author} {\bibfnamefont {P.}~\bibnamefont
  {Coleman}}, \bibinfo {author} {\bibfnamefont {L.~B.}\ \bibnamefont {Ioffe}},
  \ and\ \bibinfo {author} {\bibfnamefont {A.~M.}\ \bibnamefont {Tsvelik}},\
  }\bibfield  {title} {\enquote {\bibinfo {title} {Simple formulation of the
  two-channel kondo model},}\ }\href
  {http://link.aps.org/doi/10.1103/PhysRevB.52.6611} {\bibfield  {journal}
  {\bibinfo  {journal} {Phys. Rev. B}\ }\textbf {\bibinfo {volume} {52}},\
  \bibinfo {pages} {6611--6627} (\bibinfo {year} {1995})}\BibitemShut {NoStop}%
\bibitem [{\citenamefont {Mora}\ and\ \citenamefont {Le~Hur}(2013)}]{Mora2013}%
  \BibitemOpen
  \bibfield  {author} {\bibinfo {author} {\bibfnamefont {C.}~\bibnamefont
  {Mora}}\ and\ \bibinfo {author} {\bibfnamefont {K.}~\bibnamefont {Le~Hur}},\
  }\bibfield  {title} {\enquote {\bibinfo {title} {Probing dynamics of majorana
  fermions in quantum impurity systems},}\ }\href
  {http://link.aps.org/doi/10.1103/PhysRevB.88.241302} {\bibfield  {journal}
  {\bibinfo  {journal} {Phys. Rev. B}\ }\textbf {\bibinfo {volume} {88}},\
  \bibinfo {pages} {241302} (\bibinfo {year} {2013})}\BibitemShut {NoStop}%
\bibitem [{\citenamefont {{Iftikhar}}\ \emph {et~al.}(2015)\citenamefont
  {{Iftikhar}}, \citenamefont {{Jezouin}}, \citenamefont {{Anthore}},
  \citenamefont {{Gennser}}, \citenamefont {{Parmentier}}, \citenamefont
  {{Cavanna}},\ and\ \citenamefont {{Pierre}}}]{Iftikhar2015}%
  \BibitemOpen
  \bibfield  {author} {\bibinfo {author} {\bibfnamefont {Z.}~\bibnamefont
  {{Iftikhar}}}, \bibinfo {author} {\bibfnamefont {S.}~\bibnamefont
  {{Jezouin}}}, \bibinfo {author} {\bibfnamefont {A.}~\bibnamefont
  {{Anthore}}}, \bibinfo {author} {\bibfnamefont {U.}~\bibnamefont
  {{Gennser}}}, \bibinfo {author} {\bibfnamefont {F.~D.}\ \bibnamefont
  {{Parmentier}}}, \bibinfo {author} {\bibfnamefont {A.}~\bibnamefont
  {{Cavanna}}}, \ and\ \bibinfo {author} {\bibfnamefont {F.}~\bibnamefont
  {{Pierre}}},\ }\bibfield  {title} {\enquote {\bibinfo {title} {{Two-channel
  Kondo effect and renormalization flow with macroscopic quantum charge
  states}},}\ }\href
  {http://www.nature.com/nature/journal/v526/n7572/full/nature15384.html}
  {\bibfield  {journal} {\bibinfo  {journal} {\nat}\ }\textbf {\bibinfo
  {volume} {526}},\ \bibinfo {pages} {233--236} (\bibinfo {year}
  {2015})}\BibitemShut {NoStop}%
\bibitem [{\citenamefont {Potok}\ \emph {et~al.}(2007)\citenamefont {Potok},
  \citenamefont {Rau}, \citenamefont {Shtrikman}, \citenamefont {Oreg},\ and\
  \citenamefont {Goldhaber-Gordon}}]{potok2007}%
  \BibitemOpen
  \bibfield  {author} {\bibinfo {author} {\bibfnamefont {R.~M.}\ \bibnamefont
  {Potok}}, \bibinfo {author} {\bibfnamefont {I.~G.}\ \bibnamefont {Rau}},
  \bibinfo {author} {\bibfnamefont {H.}~\bibnamefont {Shtrikman}}, \bibinfo
  {author} {\bibfnamefont {Y.}~\bibnamefont {Oreg}}, \ and\ \bibinfo {author}
  {\bibfnamefont {D.}~\bibnamefont {Goldhaber-Gordon}},\ }\bibfield  {title}
  {\enquote {\bibinfo {title} {Observation of the two-channel kondo effect},}\
  }\href
  {http://www.nature.com/nature/journal/v446/n7132/full/nature05556.html}
  {\bibfield  {journal} {\bibinfo  {journal} {Nature}\ }\textbf {\bibinfo
  {volume} {446}},\ \bibinfo {pages} {167--171} (\bibinfo {year}
  {2007})}\BibitemShut {NoStop}%
\bibitem [{\citenamefont {{Keller}}\ \emph {et~al.}(2015)\citenamefont
  {{Keller}}, \citenamefont {{Peeters}}, \citenamefont {{Moca}}, \citenamefont
  {{Weymann}}, \citenamefont {{Mahalu}}, \citenamefont {{Umansky}},
  \citenamefont {{Zar{\'a}nd}},\ and\ \citenamefont
  {{Goldhaber-Gordon}}}]{Keller2015}%
  \BibitemOpen
  \bibfield  {author} {\bibinfo {author} {\bibfnamefont {A.~J.}\ \bibnamefont
  {{Keller}}}, \bibinfo {author} {\bibfnamefont {L.}~\bibnamefont {{Peeters}}},
  \bibinfo {author} {\bibfnamefont {C.~P.}\ \bibnamefont {{Moca}}}, \bibinfo
  {author} {\bibfnamefont {I.}~\bibnamefont {{Weymann}}}, \bibinfo {author}
  {\bibfnamefont {D.}~\bibnamefont {{Mahalu}}}, \bibinfo {author}
  {\bibfnamefont {V.}~\bibnamefont {{Umansky}}}, \bibinfo {author}
  {\bibfnamefont {G.}~\bibnamefont {{Zar{\'a}nd}}}, \ and\ \bibinfo {author}
  {\bibfnamefont {D.}~\bibnamefont {{Goldhaber-Gordon}}},\ }\bibfield  {title}
  {\enquote {\bibinfo {title} {{Universal Fermi liquid crossover and quantum
  criticality in a mesoscopic system}},}\ }\href
  {http://www.nature.com/nature/journal/v526/n7572/full/nature15261.html}
  {\bibfield  {journal} {\bibinfo  {journal} {\nat}\ }\textbf {\bibinfo
  {volume} {526}},\ \bibinfo {pages} {237--240} (\bibinfo {year}
  {2015})}\BibitemShut {NoStop}%
\bibitem [{\citenamefont {Golub}\ \emph {et~al.}(2011)\citenamefont {Golub},
  \citenamefont {Kuzmenko},\ and\ \citenamefont {Avishai}}]{golub2011}%
  \BibitemOpen
  \bibfield  {author} {\bibinfo {author} {\bibfnamefont {A.}~\bibnamefont
  {Golub}}, \bibinfo {author} {\bibfnamefont {I.}~\bibnamefont {Kuzmenko}}, \
  and\ \bibinfo {author} {\bibfnamefont {Y.}~\bibnamefont {Avishai}},\
  }\bibfield  {title} {\enquote {\bibinfo {title} {Kondo correlations and
  majorana bound states in a metal to quantum-dot to topological-superconductor
  junction},}\ }\href {\doibase 10.1103/PhysRevLett.107.176802} {\bibfield
  {journal} {\bibinfo  {journal} {Phys. Rev. Lett.}\ }\textbf {\bibinfo
  {volume} {107}},\ \bibinfo {pages} {176802} (\bibinfo {year}
  {2011})}\BibitemShut {NoStop}%
\bibitem [{\citenamefont {Lee}\ \emph {et~al.}(2013)\citenamefont {Lee},
  \citenamefont {Lim},\ and\ \citenamefont {L\'opez}}]{lee2013}%
  \BibitemOpen
  \bibfield  {author} {\bibinfo {author} {\bibfnamefont {M.}~\bibnamefont
  {Lee}}, \bibinfo {author} {\bibfnamefont {J.~S.}\ \bibnamefont {Lim}}, \ and\
  \bibinfo {author} {\bibfnamefont {R.}~\bibnamefont {L\'opez}},\ }\bibfield
  {title} {\enquote {\bibinfo {title} {Kondo effect in a quantum dot
  side-coupled to a topological superconductor},}\ }\href {\doibase
  10.1103/PhysRevB.87.241402} {\bibfield  {journal} {\bibinfo  {journal} {Phys.
  Rev. B}\ }\textbf {\bibinfo {volume} {87}},\ \bibinfo {pages} {241402}
  (\bibinfo {year} {2013})}\BibitemShut {NoStop}%
\bibitem [{\citenamefont {Cheng}\ \emph {et~al.}(2014)\citenamefont {Cheng},
  \citenamefont {Becker}, \citenamefont {Bauer},\ and\ \citenamefont
  {Lutchyn}}]{meng2014}%
  \BibitemOpen
  \bibfield  {author} {\bibinfo {author} {\bibfnamefont {M.}~\bibnamefont
  {Cheng}}, \bibinfo {author} {\bibfnamefont {M.}~\bibnamefont {Becker}},
  \bibinfo {author} {\bibfnamefont {B.}~\bibnamefont {Bauer}}, \ and\ \bibinfo
  {author} {\bibfnamefont {R.~M.}\ \bibnamefont {Lutchyn}},\ }\bibfield
  {title} {\enquote {\bibinfo {title} {Interplay between kondo and majorana
  interactions in quantum dots},}\ }\href {\doibase 10.1103/PhysRevX.4.031051}
  {\bibfield  {journal} {\bibinfo  {journal} {Phys. Rev. X}\ }\textbf {\bibinfo
  {volume} {4}},\ \bibinfo {pages} {031051} (\bibinfo {year}
  {2014})}\BibitemShut {NoStop}%
\bibitem [{\citenamefont {Fu}(2010)}]{Fu2010}%
  \BibitemOpen
  \bibfield  {author} {\bibinfo {author} {\bibfnamefont {L.}~\bibnamefont
  {Fu}},\ }\bibfield  {title} {\enquote {\bibinfo {title} {Electron
  teleportation via majorana bound states in a mesoscopic superconductor},}\
  }\href {http://link.aps.org/doi/10.1103/PhysRevLett.104.056402} {\bibfield
  {journal} {\bibinfo  {journal} {Phys. Rev. Lett.}\ }\textbf {\bibinfo
  {volume} {104}},\ \bibinfo {pages} {056402} (\bibinfo {year}
  {2010})}\BibitemShut {NoStop}%
\bibitem [{\citenamefont {Zazunov}\ \emph {et~al.}(2011)\citenamefont
  {Zazunov}, \citenamefont {Yeyati},\ and\ \citenamefont
  {Egger}}]{Zazunov2011}%
  \BibitemOpen
  \bibfield  {author} {\bibinfo {author} {\bibfnamefont {A.}~\bibnamefont
  {Zazunov}}, \bibinfo {author} {\bibfnamefont {A.~Levy}\ \bibnamefont
  {Yeyati}}, \ and\ \bibinfo {author} {\bibfnamefont {R.}~\bibnamefont
  {Egger}},\ }\bibfield  {title} {\enquote {\bibinfo {title} {Coulomb blockade
  of majorana-fermion-induced transport},}\ }\href
  {http://link.aps.org/doi/10.1103/PhysRevB.84.165440} {\bibfield  {journal}
  {\bibinfo  {journal} {Phys. Rev. B}\ }\textbf {\bibinfo {volume} {84}},\
  \bibinfo {pages} {165440} (\bibinfo {year} {2011})}\BibitemShut {NoStop}%
\bibitem [{\citenamefont {van Heck}\ \emph {et~al.}(2016)\citenamefont {van
  Heck}, \citenamefont {Lutchyn},\ and\ \citenamefont {Glazman}}]{vanHeck2016}%
  \BibitemOpen
  \bibfield  {author} {\bibinfo {author} {\bibfnamefont {B.}~\bibnamefont {van
  Heck}}, \bibinfo {author} {\bibfnamefont {R.~M.}\ \bibnamefont {Lutchyn}}, \
  and\ \bibinfo {author} {\bibfnamefont {L.~I.}\ \bibnamefont {Glazman}},\
  }\bibfield  {title} {\enquote {\bibinfo {title} {Conductance of a
  proximitized nanowire in the coulomb blockade regime},}\ }\href
  {http://link.aps.org/doi/10.1103/PhysRevB.93.235431} {\bibfield  {journal}
  {\bibinfo  {journal} {Phys. Rev. B}\ }\textbf {\bibinfo {volume} {93}},\
  \bibinfo {pages} {235431} (\bibinfo {year} {2016})}\BibitemShut {NoStop}%
\bibitem [{\citenamefont {Golub}\ and\ \citenamefont
  {Grosfeld}(2012)}]{Golub2012}%
  \BibitemOpen
  \bibfield  {author} {\bibinfo {author} {\bibfnamefont {A.}~\bibnamefont
  {Golub}}\ and\ \bibinfo {author} {\bibfnamefont {E.}~\bibnamefont
  {Grosfeld}},\ }\bibfield  {title} {\enquote {\bibinfo {title} {Charge
  resistance in a majorana $rc$ circuit},}\ }\href
  {http://link.aps.org/doi/10.1103/PhysRevB.86.241105} {\bibfield  {journal}
  {\bibinfo  {journal} {Phys. Rev. B}\ }\textbf {\bibinfo {volume} {86}},\
  \bibinfo {pages} {241105} (\bibinfo {year} {2012})}\BibitemShut {NoStop}%
\bibitem [{\citenamefont {Lee}\ and\ \citenamefont {Choi}(2014)}]{lee2014}%
  \BibitemOpen
  \bibfield  {author} {\bibinfo {author} {\bibfnamefont {M.}~\bibnamefont
  {Lee}}\ and\ \bibinfo {author} {\bibfnamefont {M.-S.}\ \bibnamefont {Choi}},\
  }\bibfield  {title} {\enquote {\bibinfo {title} {Quantum resistor-capacitor
  circuit with majorana fermion modes in a chiral topological
  superconductor},}\ }\href {\doibase 10.1103/PhysRevLett.113.076801}
  {\bibfield  {journal} {\bibinfo  {journal} {Phys. Rev. Lett.}\ }\textbf
  {\bibinfo {volume} {113}},\ \bibinfo {pages} {076801} (\bibinfo {year}
  {2014})}\BibitemShut {NoStop}%
\bibitem [{\citenamefont {B\'eri}\ and\ \citenamefont
  {Cooper}(2012)}]{Beri-2012}%
  \BibitemOpen
  \bibfield  {author} {\bibinfo {author} {\bibfnamefont {B.}~\bibnamefont
  {B\'eri}}\ and\ \bibinfo {author} {\bibfnamefont {N.~R.}\ \bibnamefont
  {Cooper}},\ }\bibfield  {title} {\enquote {\bibinfo {title} {Topological
  kondo effect with majorana fermions},}\ }\href {\doibase
  10.1103/PhysRevLett.109.156803} {\bibfield  {journal} {\bibinfo  {journal}
  {Phys. Rev. Lett.}\ }\textbf {\bibinfo {volume} {109}},\ \bibinfo {pages}
  {156803} (\bibinfo {year} {2012})}\BibitemShut {NoStop}%
\bibitem [{\citenamefont {Altland}\ and\ \citenamefont
  {Egger}(2013)}]{Atland-2013}%
  \BibitemOpen
  \bibfield  {author} {\bibinfo {author} {\bibfnamefont {A.}~\bibnamefont
  {Altland}}\ and\ \bibinfo {author} {\bibfnamefont {R.}~\bibnamefont
  {Egger}},\ }\bibfield  {title} {\enquote {\bibinfo {title} {Multiterminal
  coulomb-majorana junction},}\ }\href {\doibase
  10.1103/PhysRevLett.110.196401} {\bibfield  {journal} {\bibinfo  {journal}
  {Phys. Rev. Lett.}\ }\textbf {\bibinfo {volume} {110}},\ \bibinfo {pages}
  {196401} (\bibinfo {year} {2013})}\BibitemShut {NoStop}%
\bibitem [{\citenamefont {{Krogstrup}}\ \emph {et~al.}(2015)\citenamefont
  {{Krogstrup}}, \citenamefont {{Ziino}}, \citenamefont {{Chang}},
  \citenamefont {{Albrecht}}, \citenamefont {{Madsen}}, \citenamefont
  {{Johnson}}, \citenamefont {{Nyg{\aa}rd}}, \citenamefont {{Marcus}},\ and\
  \citenamefont {{Jespersen}}}]{Krogstrup2015}%
  \BibitemOpen
  \bibfield  {author} {\bibinfo {author} {\bibfnamefont {P.}~\bibnamefont
  {{Krogstrup}}}, \bibinfo {author} {\bibfnamefont {N.~L.~B.}\ \bibnamefont
  {{Ziino}}}, \bibinfo {author} {\bibfnamefont {W.}~\bibnamefont {{Chang}}},
  \bibinfo {author} {\bibfnamefont {S.~M.}\ \bibnamefont {{Albrecht}}},
  \bibinfo {author} {\bibfnamefont {M.~H.}\ \bibnamefont {{Madsen}}}, \bibinfo
  {author} {\bibfnamefont {E.}~\bibnamefont {{Johnson}}}, \bibinfo {author}
  {\bibfnamefont {J.}~\bibnamefont {{Nyg{\aa}rd}}}, \bibinfo {author}
  {\bibfnamefont {C.~M.}\ \bibnamefont {{Marcus}}}, \ and\ \bibinfo {author}
  {\bibfnamefont {T.~S.}\ \bibnamefont {{Jespersen}}},\ }\bibfield  {title}
  {\enquote {\bibinfo {title} {{Epitaxy of semiconductor-superconductor
  nanowires}},}\ }\href
  {http://www.nature.com/nmat/journal/v14/n4/full/nmat4176.html} {\bibfield
  {journal} {\bibinfo  {journal} {Nature Materials}\ }\textbf {\bibinfo
  {volume} {14}},\ \bibinfo {pages} {400--406} (\bibinfo {year}
  {2015})}\BibitemShut {NoStop}%
\bibitem [{\citenamefont {Buccheri}\ \emph {et~al.}(2016)\citenamefont
  {Buccheri}, \citenamefont {Bruce}, \citenamefont {Trombettoni}, \citenamefont
  {Cassettari}, \citenamefont {Babujian}, \citenamefont {Korepin},\ and\
  \citenamefont {Sodano}}]{Buccheri2016}%
  \BibitemOpen
  \bibfield  {author} {\bibinfo {author} {\bibfnamefont {F.}~\bibnamefont
  {Buccheri}}, \bibinfo {author} {\bibfnamefont {G.~D.}\ \bibnamefont {Bruce}},
  \bibinfo {author} {\bibfnamefont {A.}~\bibnamefont {Trombettoni}}, \bibinfo
  {author} {\bibfnamefont {D.}~\bibnamefont {Cassettari}}, \bibinfo {author}
  {\bibfnamefont {H.}~\bibnamefont {Babujian}}, \bibinfo {author}
  {\bibfnamefont {V.~E.}\ \bibnamefont {Korepin}}, \ and\ \bibinfo {author}
  {\bibfnamefont {P.}~\bibnamefont {Sodano}},\ }\bibfield  {title} {\enquote
  {\bibinfo {title} {Holographic optical traps for atom-based topological kondo
  devices},}\ }\href {http://stacks.iop.org/1367-2630/18/i=7/a=075012}
  {\bibfield  {journal} {\bibinfo  {journal} {New Journal of Physics}\ }\textbf
  {\bibinfo {volume} {18}},\ \bibinfo {pages} {075012} (\bibinfo {year}
  {2016})}\BibitemShut {NoStop}%
\bibitem [{\citenamefont {Eriksson}\ \emph
  {et~al.}(2014{\natexlab{a}})\citenamefont {Eriksson}, \citenamefont {Nava},
  \citenamefont {Mora},\ and\ \citenamefont {Egger}}]{Eriksson-2014}%
  \BibitemOpen
  \bibfield  {author} {\bibinfo {author} {\bibfnamefont {E.}~\bibnamefont
  {Eriksson}}, \bibinfo {author} {\bibfnamefont {A.}~\bibnamefont {Nava}},
  \bibinfo {author} {\bibfnamefont {C.}~\bibnamefont {Mora}}, \ and\ \bibinfo
  {author} {\bibfnamefont {R.}~\bibnamefont {Egger}},\ }\bibfield  {title}
  {\enquote {\bibinfo {title} {Tunneling spectroscopy of majorana-kondo
  devices},}\ }\href {\doibase 10.1103/PhysRevB.90.245417} {\bibfield
  {journal} {\bibinfo  {journal} {Phys. Rev. B}\ }\textbf {\bibinfo {volume}
  {90}},\ \bibinfo {pages} {245417} (\bibinfo {year}
  {2014}{\natexlab{a}})}\BibitemShut {NoStop}%
\bibitem [{\citenamefont {Eriksson}\ \emph
  {et~al.}(2014{\natexlab{b}})\citenamefont {Eriksson}, \citenamefont {Mora},
  \citenamefont {Zazunov},\ and\ \citenamefont
  {Egger}}]{Eriksson-2014-Manifold}%
  \BibitemOpen
  \bibfield  {author} {\bibinfo {author} {\bibfnamefont {E.}~\bibnamefont
  {Eriksson}}, \bibinfo {author} {\bibfnamefont {C.}~\bibnamefont {Mora}},
  \bibinfo {author} {\bibfnamefont {A.}~\bibnamefont {Zazunov}}, \ and\
  \bibinfo {author} {\bibfnamefont {R.}~\bibnamefont {Egger}},\ }\bibfield
  {title} {\enquote {\bibinfo {title} {Non-fermi-liquid manifold in a majorana
  device},}\ }\href {\doibase 10.1103/PhysRevLett.113.076404} {\bibfield
  {journal} {\bibinfo  {journal} {Phys. Rev. Lett.}\ }\textbf {\bibinfo
  {volume} {113}},\ \bibinfo {pages} {076404} (\bibinfo {year}
  {2014}{\natexlab{b}})}\BibitemShut {NoStop}%
\bibitem [{\citenamefont {B\'eri}(2013)}]{Beri-2013}%
  \BibitemOpen
  \bibfield  {author} {\bibinfo {author} {\bibfnamefont {B.}~\bibnamefont
  {B\'eri}},\ }\bibfield  {title} {\enquote {\bibinfo {title} {Majorana-klein
  hybridization in topological superconductor junctions},}\ }\href {\doibase
  10.1103/PhysRevLett.110.216803} {\bibfield  {journal} {\bibinfo  {journal}
  {Phys. Rev. Lett.}\ }\textbf {\bibinfo {volume} {110}},\ \bibinfo {pages}
  {216803} (\bibinfo {year} {2013})}\BibitemShut {NoStop}%
\bibitem [{\citenamefont {Galpin}\ \emph {et~al.}(2014)\citenamefont {Galpin},
  \citenamefont {Mitchell}, \citenamefont {Temaismithi}, \citenamefont {Logan},
  \citenamefont {B\'eri},\ and\ \citenamefont {Cooper}}]{Galpin-2014}%
  \BibitemOpen
  \bibfield  {author} {\bibinfo {author} {\bibfnamefont {M.~R.}\ \bibnamefont
  {Galpin}}, \bibinfo {author} {\bibfnamefont {A.~K.}\ \bibnamefont
  {Mitchell}}, \bibinfo {author} {\bibfnamefont {J.}~\bibnamefont
  {Temaismithi}}, \bibinfo {author} {\bibfnamefont {D.~E.}\ \bibnamefont
  {Logan}}, \bibinfo {author} {\bibfnamefont {B.}~\bibnamefont {B\'eri}}, \
  and\ \bibinfo {author} {\bibfnamefont {N.~R.}\ \bibnamefont {Cooper}},\
  }\bibfield  {title} {\enquote {\bibinfo {title} {Conductance fingerprint of
  majorana fermions in the topological kondo effect},}\ }\href {\doibase
  10.1103/PhysRevB.89.045143} {\bibfield  {journal} {\bibinfo  {journal} {Phys.
  Rev. B}\ }\textbf {\bibinfo {volume} {89}},\ \bibinfo {pages} {045143}
  (\bibinfo {year} {2014})}\BibitemShut {NoStop}%
\bibitem [{\citenamefont {Zazunov}\ \emph {et~al.}(2014)\citenamefont
  {Zazunov}, \citenamefont {Atland},\ and\ \citenamefont
  {Egger}}]{Zazunov-2014}%
  \BibitemOpen
  \bibfield  {author} {\bibinfo {author} {\bibfnamefont {A.}~\bibnamefont
  {Zazunov}}, \bibinfo {author} {\bibfnamefont {A.}~\bibnamefont {Atland}}, \
  and\ \bibinfo {author} {\bibfnamefont {R.}~\bibnamefont {Egger}},\ }\bibfield
   {title} {\enquote {\bibinfo {title} {Transport properties of the
  coulomb–majorana junction},}\ }\href
  {http://stacks.iop.org/1367-2630/16/i=1/a=015010} {\bibfield  {journal}
  {\bibinfo  {journal} {New Journal of Physics}\ }\textbf {\bibinfo {volume}
  {16}},\ \bibinfo {pages} {015010} (\bibinfo {year} {2014})}\BibitemShut
  {NoStop}%
\bibitem [{\citenamefont {Altland}\ \emph {et~al.}(2014)\citenamefont
  {Altland}, \citenamefont {B\'eri}, \citenamefont {Egger},\ and\ \citenamefont
  {Tsvelik}}]{Atland-2014}%
  \BibitemOpen
  \bibfield  {author} {\bibinfo {author} {\bibfnamefont {A.}~\bibnamefont
  {Altland}}, \bibinfo {author} {\bibfnamefont {B.}~\bibnamefont {B\'eri}},
  \bibinfo {author} {\bibfnamefont {R.}~\bibnamefont {Egger}}, \ and\ \bibinfo
  {author} {\bibfnamefont {A.~M.}\ \bibnamefont {Tsvelik}},\ }\bibfield
  {title} {\enquote {\bibinfo {title} {Multichannel kondo impurity dynamics in
  a majorana device},}\ }\href {\doibase 10.1103/PhysRevLett.113.076401}
  {\bibfield  {journal} {\bibinfo  {journal} {Phys. Rev. Lett.}\ }\textbf
  {\bibinfo {volume} {113}},\ \bibinfo {pages} {076401} (\bibinfo {year}
  {2014})}\BibitemShut {NoStop}%
\bibitem [{\citenamefont {Plugge}\ \emph {et~al.}(2016)\citenamefont {Plugge},
  \citenamefont {Zazunov}, \citenamefont {Eriksson}, \citenamefont {Tsvelik},\
  and\ \citenamefont {Egger}}]{Plugge-2016}%
  \BibitemOpen
  \bibfield  {author} {\bibinfo {author} {\bibfnamefont {S.}~\bibnamefont
  {Plugge}}, \bibinfo {author} {\bibfnamefont {A.}~\bibnamefont {Zazunov}},
  \bibinfo {author} {\bibfnamefont {E.}~\bibnamefont {Eriksson}}, \bibinfo
  {author} {\bibfnamefont {A.~M.}\ \bibnamefont {Tsvelik}}, \ and\ \bibinfo
  {author} {\bibfnamefont {R.}~\bibnamefont {Egger}},\ }\bibfield  {title}
  {\enquote {\bibinfo {title} {Kondo physics from quasiparticle poisoning in
  majorana devices},}\ }\href {\doibase 10.1103/PhysRevB.93.104524} {\bibfield
  {journal} {\bibinfo  {journal} {Phys. Rev. B}\ }\textbf {\bibinfo {volume}
  {93}},\ \bibinfo {pages} {104524} (\bibinfo {year} {2016})}\BibitemShut
  {NoStop}%
\bibitem [{\citenamefont {Affleck}\ and\ \citenamefont
  {Ludwig}(1993)}]{Affleck1993}%
  \BibitemOpen
  \bibfield  {author} {\bibinfo {author} {\bibfnamefont {I.}~\bibnamefont
  {Affleck}}\ and\ \bibinfo {author} {\bibfnamefont {A.~W.~W.}\ \bibnamefont
  {Ludwig}},\ }\bibfield  {title} {\enquote {\bibinfo {title} {Exact
  conformal-field-theory results on the multichannel kondo effect:
  Single-fermion green's function, self-energy, and resistivity},}\ }\href
  {http://link.aps.org/doi/10.1103/PhysRevB.48.7297} {\bibfield  {journal}
  {\bibinfo  {journal} {Phys. Rev. B}\ }\textbf {\bibinfo {volume} {48}},\
  \bibinfo {pages} {7297--7321} (\bibinfo {year} {1993})}\BibitemShut {NoStop}%
\bibitem [{\citenamefont {Yi}\ and\ \citenamefont {Kane}(1998)}]{Yi-1998}%
  \BibitemOpen
  \bibfield  {author} {\bibinfo {author} {\bibfnamefont {H.}~\bibnamefont
  {Yi}}\ and\ \bibinfo {author} {\bibfnamefont {C.~L.}\ \bibnamefont {Kane}},\
  }\bibfield  {title} {\enquote {\bibinfo {title} {Quantum brownian motion in a
  periodic potential and the multichannel kondo problem},}\ }\href {\doibase
  10.1103/PhysRevB.57.R5579} {\bibfield  {journal} {\bibinfo  {journal} {Phys.
  Rev. B}\ }\textbf {\bibinfo {volume} {57}},\ \bibinfo {pages} {R5579--R5582}
  (\bibinfo {year} {1998})}\BibitemShut {NoStop}%
\bibitem [{\citenamefont {Yi}(2002)}]{Yi-2002}%
  \BibitemOpen
  \bibfield  {author} {\bibinfo {author} {\bibfnamefont {H.}~\bibnamefont
  {Yi}},\ }\bibfield  {title} {\enquote {\bibinfo {title} {Resonant tunneling
  and the multichannel kondo problem: Quantum brownian motion description},}\
  }\href {\doibase 10.1103/PhysRevB.65.195101} {\bibfield  {journal} {\bibinfo
  {journal} {Phys. Rev. B}\ }\textbf {\bibinfo {volume} {65}},\ \bibinfo
  {pages} {195101} (\bibinfo {year} {2002})}\BibitemShut {NoStop}%
\bibitem [{\citenamefont {{Michaeli}}\ \emph {et~al.}(2016)\citenamefont
  {{Michaeli}}, \citenamefont {{Aviad Landau}}, \citenamefont {{Sela}},\ and\
  \citenamefont {{Fu}}}]{Michaeli2016}%
  \BibitemOpen
  \bibfield  {author} {\bibinfo {author} {\bibfnamefont {K.}~\bibnamefont
  {{Michaeli}}}, \bibinfo {author} {\bibfnamefont {L.}~\bibnamefont {{Aviad
  Landau}}}, \bibinfo {author} {\bibfnamefont {E.}~\bibnamefont {{Sela}}}, \
  and\ \bibinfo {author} {\bibfnamefont {L.}~\bibnamefont {{Fu}}},\ }\bibfield
  {title} {\enquote {\bibinfo {title} {{Electron Teleportation in
  Multi-Terminal Majorana Islands: Statistical Transmutation and Fractional
  Quantum Conductance}},}\ }\href {http://arxiv.org/abs/1608.00581} {\bibfield
  {journal} {\bibinfo  {journal} {ArXiv e-prints}\ } (\bibinfo {year}
  {2016})},\ \Eprint {http://arxiv.org/abs/1608.00581} {arXiv:1608.00581
  [cond-mat.mes-hall]} \BibitemShut {NoStop}%
\bibitem [{\citenamefont {Haldane}(1981)}]{Haldane1981}%
  \BibitemOpen
  \bibfield  {author} {\bibinfo {author} {\bibfnamefont {F.~D.~M.}\
  \bibnamefont {Haldane}},\ }\bibfield  {title} {\enquote {\bibinfo {title}
  {Luttinger liquid theory of one-dimensional quantum fluids. i. properties of
  the luttinger model and their extension to the general 1d interacting
  spinless fermi gas},}\ }\href
  {http://iopscience.iop.org/article/10.1088/0022-3719/14/19/010/meta}
  {\bibfield  {journal} {\bibinfo  {journal} {J. Phys. C 14 2585}\ } (\bibinfo
  {year} {1981})}\BibitemShut {NoStop}%
\bibitem [{\citenamefont {Giamarchi}(2004)}]{QP1DGiamarchi2004}%
  \BibitemOpen
  \bibfield  {author} {\bibinfo {author} {\bibfnamefont {T.}~\bibnamefont
  {Giamarchi}},\ }\href@noop {} {\emph {\bibinfo {title} {Quantum Physics in
  One Dimension}}},\ edited by\ \bibinfo {editor} {\bibfnamefont
  {Oxford~University}\ \bibnamefont {Press}}\ (\bibinfo {year}
  {2004})\BibitemShut {NoStop}%
\bibitem [{\citenamefont {Gogolin}\ \emph {et~al.}(1998)\citenamefont
  {Gogolin}, \citenamefont {Nersesyan},\ and\ \citenamefont
  {Tsvelik}}]{BosonizationGogolin1998}%
  \BibitemOpen
  \bibfield  {author} {\bibinfo {author} {\bibfnamefont {A.O.}\ \bibnamefont
  {Gogolin}}, \bibinfo {author} {\bibfnamefont {A.A.}\ \bibnamefont
  {Nersesyan}}, \ and\ \bibinfo {author} {\bibfnamefont {A.M.}\ \bibnamefont
  {Tsvelik}},\ }\href@noop {} {\emph {\bibinfo {title} {Bosonization and
  strongly correlated systems}}},\ edited by\ \bibinfo {editor} {\bibfnamefont
  {Cambridge~University}\ \bibnamefont {Press}}\ (\bibinfo {year}
  {1998})\BibitemShut {NoStop}%
\bibitem [{\citenamefont {Schmid}(1983)}]{schmid1983}%
  \BibitemOpen
  \bibfield  {author} {\bibinfo {author} {\bibfnamefont {A.}~\bibnamefont
  {Schmid}},\ }\bibfield  {title} {\enquote {\bibinfo {title} {Diffusion and
  localization in a dissipative quantum system},}\ }\href {\doibase
  10.1103/PhysRevLett.51.1506} {\bibfield  {journal} {\bibinfo  {journal}
  {Phys. Rev. Lett.}\ }\textbf {\bibinfo {volume} {51}},\ \bibinfo {pages}
  {1506--1509} (\bibinfo {year} {1983})}\BibitemShut {NoStop}%
\bibitem [{\citenamefont {Kane}\ and\ \citenamefont {Fisher}(1992)}]{Kane1992}%
  \BibitemOpen
  \bibfield  {author} {\bibinfo {author} {\bibfnamefont {C.~L.}\ \bibnamefont
  {Kane}}\ and\ \bibinfo {author} {\bibfnamefont {Matthew P.~A.}\ \bibnamefont
  {Fisher}},\ }\bibfield  {title} {\enquote {\bibinfo {title} {Transmission
  through barriers and resonant tunneling in an interacting one-dimensional
  electron gas},}\ }\href {http://link.aps.org/doi/10.1103/PhysRevB.46.15233}
  {\bibfield  {journal} {\bibinfo  {journal} {Phys. Rev. B}\ }\textbf {\bibinfo
  {volume} {46}},\ \bibinfo {pages} {15233--15262} (\bibinfo {year}
  {1992})}\BibitemShut {NoStop}%
\bibitem [{\citenamefont {Anderson}(1970)}]{Anderson1970-poorman}%
  \BibitemOpen
  \bibfield  {author} {\bibinfo {author} {\bibfnamefont {P.~W.}\ \bibnamefont
  {Anderson}},\ }\bibfield  {title} {\enquote {\bibinfo {title} {A poor man's
  derivation of scaling laws for the kondo problem},}\ }\href
  {http://stacks.iop.org/0022-3719/3/i=12/a=008} {\bibfield  {journal}
  {\bibinfo  {journal} {Journal of Physics C: Solid State Physics}\ }\textbf
  {\bibinfo {volume} {3}},\ \bibinfo {pages} {2436} (\bibinfo {year}
  {1970})}\BibitemShut {NoStop}%
\bibitem [{\citenamefont {Safi}\ and\ \citenamefont {Schulz}(1995)}]{Safi1995}%
  \BibitemOpen
  \bibfield  {author} {\bibinfo {author} {\bibfnamefont {I.}~\bibnamefont
  {Safi}}\ and\ \bibinfo {author} {\bibfnamefont {H.~J.}\ \bibnamefont
  {Schulz}},\ }\bibfield  {title} {\enquote {\bibinfo {title} {Transport in an
  inhomogeneous interacting one-dimensional system},}\ }\href
  {http://link.aps.org/doi/10.1103/PhysRevB.52.R17040} {\bibfield  {journal}
  {\bibinfo  {journal} {Phys. Rev. B}\ }\textbf {\bibinfo {volume} {52}},\
  \bibinfo {pages} {R17040--R17043} (\bibinfo {year} {1995})}\BibitemShut
  {NoStop}%
\bibitem [{\citenamefont {Maslov}\ and\ \citenamefont
  {Stone}(1995)}]{Maslov1995}%
  \BibitemOpen
  \bibfield  {author} {\bibinfo {author} {\bibfnamefont {D.~L.}\ \bibnamefont
  {Maslov}}\ and\ \bibinfo {author} {\bibfnamefont {M.}~\bibnamefont {Stone}},\
  }\bibfield  {title} {\enquote {\bibinfo {title} {Landauer conductance of
  luttinger liquids with leads},}\ }\href
  {http://link.aps.org/doi/10.1103/PhysRevB.52.R5539} {\bibfield  {journal}
  {\bibinfo  {journal} {Phys. Rev. B}\ }\textbf {\bibinfo {volume} {52}},\
  \bibinfo {pages} {R5539--R5542} (\bibinfo {year} {1995})}\BibitemShut
  {NoStop}%
\bibitem [{\citenamefont {Anderson}\ \emph {et~al.}(1970)\citenamefont
  {Anderson}, \citenamefont {Yuval},\ and\ \citenamefont
  {Hamann}}]{Anderson1970}%
  \BibitemOpen
  \bibfield  {author} {\bibinfo {author} {\bibfnamefont {P.~W.}\ \bibnamefont
  {Anderson}}, \bibinfo {author} {\bibfnamefont {G.}~\bibnamefont {Yuval}}, \
  and\ \bibinfo {author} {\bibfnamefont {D.~R.}\ \bibnamefont {Hamann}},\
  }\bibfield  {title} {\enquote {\bibinfo {title} {Exact results in the kondo
  problem. ii. scaling theory, qualitatively correct solution, and some new
  results on one-dimensional classical statistical models},}\ }\href
  {http://link.aps.org/doi/10.1103/PhysRevB.1.4464} {\bibfield  {journal}
  {\bibinfo  {journal} {Phys. Rev. B}\ }\textbf {\bibinfo {volume} {1}},\
  \bibinfo {pages} {4464--4473} (\bibinfo {year} {1970})}\BibitemShut {NoStop}%
\bibitem [{\citenamefont {Toulouse}(1970)}]{Toulouse1970}%
  \BibitemOpen
  \bibfield  {author} {\bibinfo {author} {\bibfnamefont {G.}~\bibnamefont
  {Toulouse}},\ }\bibfield  {title} {\enquote {\bibinfo {title} {Infinite-$u$
  anderson hamiltonian for dilute alloys},}\ }\href
  {http://link.aps.org/doi/10.1103/PhysRevB.2.270} {\bibfield  {journal}
  {\bibinfo  {journal} {Phys. Rev. B}\ }\textbf {\bibinfo {volume} {2}},\
  \bibinfo {pages} {270--277} (\bibinfo {year} {1970})}\BibitemShut {NoStop}%
\bibitem [{\citenamefont {Matveev}(1994)}]{Matveev1991}%
  \BibitemOpen
  \bibfield  {author} {\bibinfo {author} {\bibfnamefont {K.~A.}\ \bibnamefont
  {Matveev}},\ }\bibfield  {title} {\enquote {\bibinfo {title} {Charge
  fluctuations under the coulomb blockade conditions},}\ }\href
  {http://www.sciencedirect.com/science/article/pii/0921452694900884}
  {\bibfield  {journal} {\bibinfo  {journal} {Physica B}\ }\textbf {\bibinfo
  {volume} {203 3-4}},\ \bibinfo {pages} {pp. 404--408} (\bibinfo {year}
  {1994})}\BibitemShut {NoStop}%
\bibitem [{\citenamefont {{Affleck}}(1995)}]{Affleck1995}%
  \BibitemOpen
  \bibfield  {author} {\bibinfo {author} {\bibfnamefont {I.}~\bibnamefont
  {{Affleck}}},\ }\bibfield  {title} {\enquote {\bibinfo {title} {{Conformal
  Field Theory Approach to the Kondo Effect}},}\ }\href
  {http://arxiv.org/abs/cond-mat/9512099} {\bibfield  {journal} {\bibinfo
  {journal} {eprint arXiv:cond-mat/9512099}\ } (\bibinfo {year} {1995})},\
  \Eprint {http://arxiv.org/abs/cond-mat/9512099} {cond-mat/9512099}
  \BibitemShut {NoStop}%
\bibitem [{\citenamefont {{Affleck}}\ \emph {et~al.}(2001)\citenamefont
  {{Affleck}}, \citenamefont {{Oshikawa}},\ and\ \citenamefont
  {{Saleur}}}]{Affleck-2001}%
  \BibitemOpen
  \bibfield  {author} {\bibinfo {author} {\bibfnamefont {I.}~\bibnamefont
  {{Affleck}}}, \bibinfo {author} {\bibfnamefont {M.}~\bibnamefont
  {{Oshikawa}}}, \ and\ \bibinfo {author} {\bibfnamefont {H.}~\bibnamefont
  {{Saleur}}},\ }\bibfield  {title} {\enquote {\bibinfo {title} {{Quantum
  brownian motion on a triangular lattice and /c=2 boundary conformal field
  theory}},}\ }\href {\doibase 10.1016/S0550-3213(00)00499-5} {\bibfield
  {journal} {\bibinfo  {journal} {Nuclear Physics B}\ }\textbf {\bibinfo
  {volume} {594}},\ \bibinfo {pages} {535--606} (\bibinfo {year}
  {2001})}\BibitemShut {NoStop}%
\bibitem [{Note1()}]{Note1}%
  \BibitemOpen
  \bibinfo {note} {The $-1$ factor was also interpreted as an effective and
  alternating $\pm \pi $ flux threading each plaquette of the triangular
  lattice~\cite {Yi-2002}.}\BibitemShut {Stop}%
\bibitem [{Note2()}]{Note2}%
  \BibitemOpen
  \bibinfo {note} {Nothing prevents the curve at charge degeneracy to cross
  above the topological Kondo line (non-degenerate case) for even lower
  mobility before turning over.}\BibitemShut {Stop}%
\bibitem [{\citenamefont {Affleck}\ \emph {et~al.}(1992)\citenamefont
  {Affleck}, \citenamefont {Ludwig}, \citenamefont {Pang},\ and\ \citenamefont
  {Cox}}]{Affleck1992}%
  \BibitemOpen
  \bibfield  {author} {\bibinfo {author} {\bibfnamefont {I.}~\bibnamefont
  {Affleck}}, \bibinfo {author} {\bibfnamefont {A.~W.~W.}\ \bibnamefont
  {Ludwig}}, \bibinfo {author} {\bibfnamefont {H.-B.}\ \bibnamefont {Pang}}, \
  and\ \bibinfo {author} {\bibfnamefont {D.~L.}\ \bibnamefont {Cox}},\
  }\bibfield  {title} {\enquote {\bibinfo {title} {Relevance of anisotropy in
  the multichannel kondo effect: Comparison of conformal field theory and
  numerical renormalization-group results},}\ }\href {\doibase
  10.1103/PhysRevB.45.7918} {\bibfield  {journal} {\bibinfo  {journal} {Phys.
  Rev. B}\ }\textbf {\bibinfo {volume} {45}},\ \bibinfo {pages} {7918--7935}
  (\bibinfo {year} {1992})}\BibitemShut {NoStop}%
\bibitem [{\citenamefont {{Zarand}}\ \emph {et~al.}(2000)\citenamefont
  {{Zarand}}, \citenamefont {{Zimanyi}},\ and\ \citenamefont
  {{Wilhelm}}}]{Zarand2000}%
  \BibitemOpen
  \bibfield  {author} {\bibinfo {author} {\bibfnamefont {G.}~\bibnamefont
  {{Zarand}}}, \bibinfo {author} {\bibfnamefont {G.~T.}\ \bibnamefont
  {{Zimanyi}}}, \ and\ \bibinfo {author} {\bibfnamefont {F.}~\bibnamefont
  {{Wilhelm}}},\ }\bibfield  {title} {\enquote {\bibinfo {title} {{Is the
  Multichannel Kondo Model Appropriate to Describe the Single Electron
  Transistor?}}}\ }\href {http://arxiv.org/abs/cond-mat/0003013} {\bibfield
  {journal} {\bibinfo  {journal} {eprint arXiv:cond-mat/0003013}\ } (\bibinfo
  {year} {2000})}\BibitemShut {NoStop}%
\bibitem [{\citenamefont {Le~Hur}\ and\ \citenamefont
  {Seelig}(2002)}]{lehur2002}%
  \BibitemOpen
  \bibfield  {author} {\bibinfo {author} {\bibfnamefont {K.}~\bibnamefont
  {Le~Hur}}\ and\ \bibinfo {author} {\bibfnamefont {G.}~\bibnamefont
  {Seelig}},\ }\bibfield  {title} {\enquote {\bibinfo {title} {Capacitance of a
  quantum dot from the channel-anisotropic two-channel kondo model},}\ }\href
  {\doibase 10.1103/PhysRevB.65.165338} {\bibfield  {journal} {\bibinfo
  {journal} {Phys. Rev. B}\ }\textbf {\bibinfo {volume} {65}},\ \bibinfo
  {pages} {165338} (\bibinfo {year} {2002})}\BibitemShut {NoStop}%
\end{thebibliography}%

\end{document}